\documentclass[aip, amsmath,amssymb, reprint]{revtex4-1}

\usepackage{natbib} 

\usepackage{amsmath}
\usepackage{epsfig}
\usepackage{array}
\usepackage[hyperfootnotes=true]{hyperref} 
\usepackage{afterpage}
\usepackage{color}
\definecolor{orange}{cmyk}{0,0.5,1,0}

\newcommand{\blue}{}

\begin{document}

\title{{\blue Code-to-Code Comparison and
Validation of the Radiation-Hydrodynamics Capabilities of the FLASH Code Using a Laboratory Astrophysical Jet}}

\author{Chris Orban}
\email{orban@physics.osu.edu}
\affiliation{Department of Physics, The Ohio State University, Columbus, OH, 43210, USA}
\author{Milad Fatenejad}
\affiliation{6157 N Kenmore, Chicago, IL 60660 }
\author{Donald Q. Lamb}
\affiliation{Department of Astronomy, University of Chicago, Chicago, IL}

\date{\today}

\begin{abstract}
The potential for laser-produced plasmas to yield fundamental insights into high energy density physics (HEDP) and deliver other useful applications can sometimes be frustrated by uncertainties in modeling the properties and behavior of these plasmas using radiation-hydrodynamics codes. In an effort to overcome this and to corroborate the accuracy of the HEDP capabilities  in the publicly available FLASH radiation-hydrodynamics code, we present detailed  code-to-code comparisons between FLASH and the HYDRA code developed at Lawrence Livermore National Laboratory using previously published HYDRA simulations from Grava et al. (2008). That study describes a laser experiment that produced a jet-like feature that the authors compare to astrophysical jets. Importantly, the Grava et al. (2008) experiment included detailed x-ray interferometric measurements of electron number densities
{\blue and a time-integrated measurement of the soft x-ray spectrum}. Despite {\blue markedly} different  methods for treating the computational mesh, and different equation of state and opacity models,  the FLASH results resemble the results from HYDRA and, most importantly, the experimental measurements of electron density. Having validated the FLASH code in this way, we use the code to further investigate and understand the formation of the jet seen in the Grava et al. (2008) experiment and discuss its relation to the Wan et al. (1997) experiment at the NOVA laser. \\
\end{abstract}


\maketitle

\defcitealias{Grava_etal2008}{GRAVA} 	
\defcitealias{Wan_etal1997}{WAN}

\section{Introduction}
  \label{sec:intro}

The potential for laser experiments to yield fundamental insights into High-Energy-Density Physics (HEDP) is in many ways limited by the sophistication and accuracy of current-generation  ``three-temperature'' (3T)\footnote{ We use the term ``three-temperature'' (or ``3T'') to denote the approximation that electrons and ions move together as a single fluid but with two different temperatures, and that this fluid can emit or absorb radiation. In the 3T simulations presented in this paper, each cell has an electron temperature, an ion temperature, and radiation energy densities in a number of photon energy bins.} radiation-hydrodynamics codes that simulate the heating, conduction and radiation of laser-irradiated fluids. In deconstructing the results from ultra-high intensity, short-pulse laser experiments, for example, Particle-In-Cell (PIC) simulations of the ultra-intense pulse interaction with the target may depend sensitively on radiation-hydrodynamics simulations of the heating and
ionizing effect of stray ``pre-pulse'' laser energy in the nanoseconds before the arrival of the main pulse. It is not always possible to use interferometric instruments to measure electron number densities in the ``pre-plasma'' created by this pre-pulse, as the target geometry may not permit probe beams to access the pre-plasma, e.g., in cone targets as in Ref.\cite{AkliOrban_etal2012}, so the pre-plasma properties must be predicted  using a radiation-hydrodynamics code. The uncertainties in these simulations may frustrate efforts to gain a better understanding of ion acceleration or electron transport that could prove to be valuable for a variety of applications, such as radiation therapy, x-ray generation, or the activation and detection of fissile materials.

Another important use of these codes is in modeling inertial confinement fusion experiments at  laser facilities  like Omega and the National Ignition Facility (NIF) \cite{Boehly_etal1997,Moses_etal2009,Lindl_etal2011}. Rosen et al. \cite{Rosen_etal2011} 
describe some of the subtleties encountered in understanding indirect-drive experiments and a 2012 panel report by Lamb \& Marinak et al. \cite{LambMarinak2012} outlines a number of remaining uncertainties in simulating ignition-relevant 
experiments  at NIF. Lamb \& Marinak et al. \cite{LambMarinak2012} emphasize the need for code-to-code comparisons and validation in a wider effort to reproduce the diagnostics of NIF implosions. Although there have been some recent investigations
with other codes \cite{Bellei_etal2013,RAGE}, the HYDRA code  
\cite{Marinak_etal1996,Marinak_etal1998,Marinak_etal2001}, developed at Lawrence
Livermore National Laboratory (LLNL), is frequently used for radiation-hydrodynamics modeling of these experiments.

Uncertainties and inaccuracies in radiation-hydrodynamics modeling can also frustrate the design and interpretation of experiments to investigate fundamental plasma properties (e.g., opacities, equation of state, hydrodynamic instabilities) at HEDP-relevant densities and temperatures  \cite{Harding_etal2009,vanderHolst_etal2012,Taccetti_etal2012}. 
Both Omega and NIF, among other facilities, have completed a number of experiments in this category, and will continue to do so in the future \cite{NIF,Keiter_etal2013}. 

With these concerns in mind, and in an effort to confirm the accuracy of the HEDP capabilities of the FLASH radiation-hydrodynamics code \cite{Fryxell_etal2000,Dubey_etal2009,Tzeferacos2014}, we compare the predictions of FLASH to previously-published results from nanosecond laser irradiation of an Aluminum target and previously published modeling of this experiment using the HYDRA code \cite{Grava_etal2008}. FLASH is a finite-volume Eulerian code that operates on a block-structured mesh using Adaptive Mesh Refinement (AMR) \cite{PARAMESH}, whereas the HYDRA code uses an Arbitrary-Lagrangian-Eulerian (ALE) scheme to determine the computational grid \cite{Hirt_etal1974,Castor2004,Kucharik2006}, which can deform and stretch in response to the movement and heating of the fluid. In other respects the codes are very similar in that they use a tabulated Equation of State (EOS) and make many of the same assumptions regarding laser propagation and absorption as well as various aspects of the hydrodynamics.

We show results for an experiment in which a target consisting of an Al slab with a mm-long triangular groove is irradiated by a rectangular laser beam. The results of this experiment, which is translationally invariant along the groove and so is a test of plasma expansion in 2D Cartesian geometry, were modeled with HYDRA simulations in Grava et al. (2008) \cite{Grava_etal2008}, hereafter referred to as GRAVA. They investigated this problem for its resemblance to astrophysical jets where radiative cooling plays an important dynamical role, and as a miniature version of similarly-motivated experiments at the Nova laser carried out by  \citet{Wan_etal1997} (hereafter WAN) and discussed by \citet{Stone_etal2000}. The experiment was performed at Colorado State University and, importantly, \citetalias{Grava_etal2008} present x-ray interferometric measurements of electron number density from 1-20 ns after the target begins to be irradiated {\blue and a time-integrated measurement of the soft x-ray spectrum} that afford {\blue useful} validation tests. We therefore compare the results of FLASH  simulations to both the HYDRA simulations  they present  and the experimental data \citetalias{Grava_etal2008} present.  

Section~\ref{sec:grava} presents comparisons of FLASH predictions to the experiments and modeling in \citetalias{Grava_etal2008}.
Section~\ref{sec:jet} uses FLASH simulations to investigate and understand the formation of the jet in the \citetalias{Grava_etal2008} experiment. We summarize and conclude in Sec.~\ref{sec:concl}.

{\blue
\section{Comparison of Radiation Hydrodynamics in the FLASH and HYDRA Codes}
\label{sec:comparison-flash-hydra}

By design the FLASH and HYDRA codes have radically different approaches to defining and evolving the computational mesh and solving the equations of hydrodynamics. 
Whereas FLASH performs 
its hydrodynamic calculations on a fixed, finite volume Eulerian grid that can be refined (or de-refined) on the fly to maintain high
resolution in interesting areas using AMR, by contrast the 
ALE approach used by HYDRA allows the grid to distort and move
with the fluid flow with (preferably) minor deviations from this Lagrangian behavior to prevent severe
tangling of the mesh.

The FLASH simulations in this paper were performed with FLASH 4-beta \cite{Flash2012} with an added Lee-More conductivity and thermal equilibration model \cite{LeeMore1984}. 
The Lee-More model became part of the publicly available version of the code starting with FLASH 4.2.2 \cite{flash_versions}.
The HYDRA simulations were previously published in \citetalias{Grava_etal2008}.  These simulations also use the Lee-More conductivity and thermal equilibration model. In both HYDRA and FLASH the electron flux limiter parameter was set to 0.05.

The FLASH and HYDRA codes differ in the EOS models they use. 
The FLASH simulations we present use the commercially-available PROPACEOS model \cite{Macfarlane_etal2006}, which is QEOS-based. The HYDRA simulations we discuss use the QEOS \cite{QEOS} model.
The EOS models give not only the pressure of the
plasma as a function of density and temperature, but also the mean ionization fraction, $\bar{Z}$, and the specific heat, which affect the electron heat conduction.

The FLASH and HYDRA simulations also use different opacity models for Al. 
The FLASH simulations use the PROPACEOS model, while \citetalias{Grava_etal2008} do not specify the opacity model used in the HYDRA simulations they presented, saying only ``All opacities were computed assuming local thermodynamic equilibrium (LTE)".
The opacity models from different research groups can disagree at the level of a few to tens of percent. 
Our own comparisons of the all-frequency-integrated Rosseland means for Al from PROPACEOS and publicly-available Opacity Project data \cite{Seaton_etal1994,Mendoza_etal2007} are consistent with this conclusion.

FLASH and HYDRA both use the Kaiser \cite{Kaiser2000} algorithm to calculate the trajectories of the laser rays. They both deposit energy into the electrons at a rate corresponding to the inverse bremsstrahlung process, and interpolate the deposited energy to the computational mesh at each timestep.

The FLASH and HYDRA simulations use very low-density Helium as a stand-in for vacuum, unless noted otherwise. 
In the problems we consider here, some of the kinetic energy of the high-velocity Al ablating from the target is transferred to the He.
Since the density of the He is very low, the temperature of the resulting  shock 
\cite{Shafranov1957,MihalasMihalas1984,Fatenejad_etal2011}
is very high. 
There may be differences in the way that FLASH and HYDRA treat the Al/He interface. However, the properties of the expanding Al plasma upstream of the interface are, in practice, insensitive to the treatment of the interface and the specific value of the density of the He because the upstream flow is supersonic and the density of the He is very low.
}

\section{Comparison to Grava et al. 2008}
\label{sec:grava}

In this section, we describe code-to-code comparisons between FLASH and HYDRA for an experiment that was carried out at Colorado State University and modeled using HYDRA  (\citetalias{Grava_etal2008}).
The \citetalias{Grava_etal2008} study is unique in both the quality of the experimental data that was collected and the sophistication of the radiation-hydrodynamics modeling that was done. In the experiment, an Al target with a V-shaped groove was irradiated by a rectangular laser beam  striking the target perpendicular to its face. The intensity of the laser beam had a Gaussian cross section with a FWHM of 360~$\mu$m in the narrow direction and was highly uniform in the wide direction. The peak of the Gaussian was aligned with the center-line of the V-shaped groove and the peak intensity of the Gaussian was $\sim 10^{12}$ W/cm$^2$. The energy of the laser was 0.8~J and the duration of the laser pulse was 120~ps. The data from the experiment was used as a validation test for HYDRA. Here we will use the data as a validation test for FLASH.

The geometry of the groove-shaped target, which is reminiscent of a well-known validation experiment done at the NOVA laser by \citetalias{Wan_etal1997} and discussed by \citet{Stone_etal2000}, allows interferometric measurements of the electron density in the blowoff plasma. \citetalias{Grava_etal2008} conducted these measurements with a few-ns cadence using soft x-rays with a wavelength of 46.9~nm. This implies a critical density of $5 \times 10^{23}$~cm$^{-3}$; however, taking into account instrumental resolution and other details, the largest measurable electron density is reported to be $5 \times 10^{20}$~cm$^{-3}$.

\citetalias{Grava_etal2008} pursued the experiment as a scaled version of astrophysical 
radiative shocks, explaining that the radiative energy loss timescale in the problem, $\tau_{\rm rad}$, is  comparable to the hydrodynamic expansion timescale, $\tau_{\rm hydro}$. 
They were also motivated by the fact that similar, earlier NOVA experiments 
produced puzzling results, raising the question whether radiation-hydrodynamics codes might be inadequate to model the experiment and collisionless Particle-In-Cell (PIC) 
codes might be needed instead(\citetalias{Wan_etal1997}). \citetalias{Grava_etal2008} and later work by the same collaboration \cite{Filevich_etal2009,Purvis_etal2010} showed that radiation-hydrodynamic codes are able to model this kind of experiment, and for a variety of different target geometries. 

\begin{figure*}
\centerline{\includegraphics[width=7in]{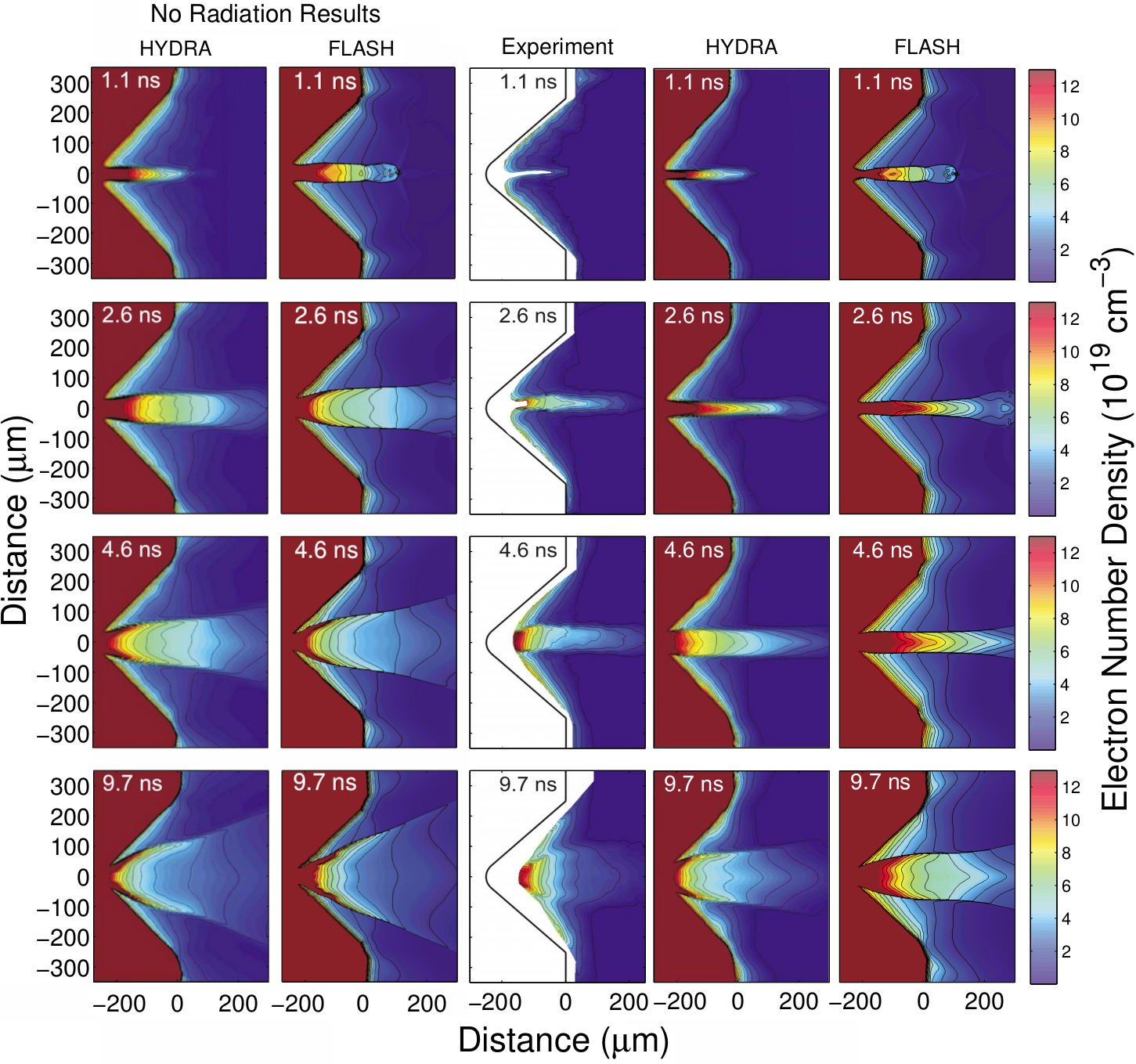}}
\caption{{\blue Left panels: Electron number densities at various times from HYDRA and FLASH simulations of the irradiation of the V-shaped groove target described in \citetalias{Grava_etal2008} that do not include radiation diffusion.  
Right panels: Comparison of the electron number density inferred from soft x-ray interferometry of the experiment described in  GRAVA\cite{Grava_etal2008} to the electron number density in HYDRA and FLASH simulations of the experiment that include multi-group radiation diffusion. 
HYDRA and Experiment panels are adapted with permission from Fig.~9 in GRAVA \cite{Grava_etal2008} (copyrighted by the American Physical Society).}
} 
\label{fig:combined}
\end{figure*}

\subsection{\,  Non-Radiative Results: \\ Electron Number Density}


\citetalias{Grava_etal2008} presents the results of HYDRA simulations with and without multi-group radiation diffusion to demonstrate the importance of radiation on their simulation results. We use \citetalias{Grava_etal2008}'s non-radiative HYDRA simulations as the starting point for our comparison, since it removes any dependence on the opacity model. We performed {\blue a} FLASH {\blue simulation} using the PROPACEOS EOS model \cite{Macfarlane_etal2006} which included physics from the so-called QEOS model \cite{QEOS} for near-solid-density interactions. {\blue The HYDRA simulations from \citetalias{Grava_etal2008} also use the QEOS model for the equation of state.}
{\blue The two left columns of Fig.~\ref{fig:combined}} compare the results of the FLASH simulation and the HYDRA simulation {\blue at different times}. The comparison shows that the results of the two simulations agree qualitatively. Specifically, there are no important features in the FLASH simulations that are not in the HYDRA simulations, and vice versa; and many of the contours from the simulations resemble each other. 

{\blue The two left columns of Fig.~\ref{fig:combined} show that the jet extends to a larger distance and is wider at earlier times in the FLASH simulations than in the HYDRA simulations.  Since radiation diffusion is not a factor, the treatment of irradiation by the laser beam is identical, and we do not expect the markedly different mesh schemes in the two codes to have a large effect in this experiment (since, e.g., turbulence is not a significant factor), we infer that the differences between the HYDRA and FLASH simulations are likely due to the modestly different EOS models in them.
}

\subsection{Results including Radiation:  \\ Electron Number Density}
\label{sec:rad}

 
Radiating plasmas can be compressed to higher densities than non-radiating plasmas because the loss of energy cools the plasma, lowering the pressure. This is evident in comparing the {\blue left-most two columns of Fig.~\ref{fig:combined}} which 
shows the spatial distribution of the electron number density at four different times for HYDRA and FLASH simulations without radiation transport, and 
{\blue the right-most two columns of Fig.~\ref{fig:combined}} which shows the same information for HYDRA and FLASH simulations with radiation transport. {\blue The center column in Fig.~\ref{fig:combined} shows} the spatial distribution of the electron number density from experimental measurements. In both
the {\blue radiative and non-radiative results, as well as the experiment, the ablating plasma collides with itself by 1.1~ns,} creating a relatively thin jet of high density, high temperature Al extending from center of the groove in the target. In {\blue the non-radiative results} the jet expands due to the high pressure, thus creating the double horn feature seen there at later times. However {\blue in the radiative results} the jet stays compressed for longer so that at 4.6~ns the width of the jet is only modestly larger. Only by 9.7~ns has ablation from the diagonal walls of the target lessened enough that the pressure {\blue in the jet can broaden it} and produce the double horn structure in the density similar to what is seen in {\blue the non-radiative results}. For a more extreme example of the effect of radiation in this problem see the Cu or Mo results in \cite{Purvis_etal2010}.

{\blue
The electron number densities in the FLASH simulations are also larger than those in the HYDRA simulations when radiation diffusion is included. The jet in the FLASH simulation is again wider and has a higher velocity through 2.6 ns.  After this, the jet in the FLASH simulations is narrower, which we attribute to greater radiative cooling in the FLASH simulation as a result of the higher degree of ionization.
 
Despite the possibility of important differences due to differences in the implementation of flux-limited multigroup diffusion and/or the opacity models, the FLASH and HYDRA simulations with radiation diffusion agree about as well as do the non-radiative simulations. Importantly, both codes agree well with the experimental data.}

\begin{figure}
\centerline{\includegraphics[angle=0,width=3.5in]{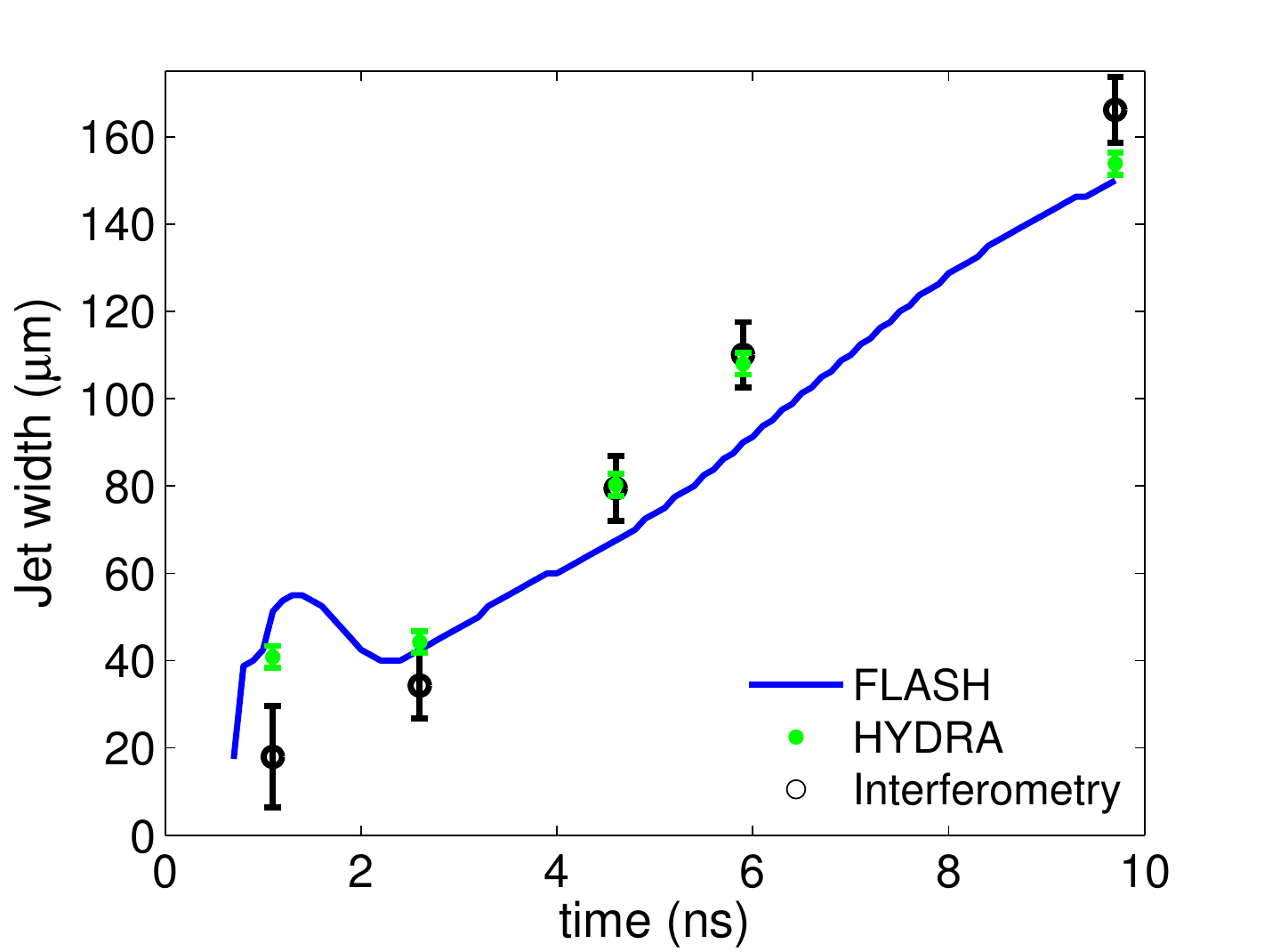}}
\vspace{-0.25cm}
\caption{A quantitative comparison of the jet width versus time as inferred from the electron number density (Fig.~\ref{fig:combined})
on a line perpendicular to the laser axis that is centered at (0,0). HYDRA results (green points) and interferometric measurements (black points)
from \citetalias{Grava_etal2008} are presented at 1.1, 2.6, 4.6, 5.9 and 9.7 ns. Measurements from the FLASH simulation (solid blue line) are more 
finely spaced in time. For brevity, the comparison at 5.9~ns is not shown in Fig.~\ref{fig:combined}.
}\label{fig:jetwidth}
\end{figure}

To provide a quantitative comparison, we compare in Fig.~\ref{fig:jetwidth} the width of the jet as a function of time as measured in the experiment and given by the FLASH and HYDRA simulations.As the width of the jet, we take the distance between the most steeply rising features in the electron density along a line perpendicular to the laser axis centered at (0,0). This definition is convenient for measuring the jet width from the figures in \citetalias{Grava_etal2008}, since the point of steepest rise is simply where the electron number density contours are closest together. We assign to the HYDRA results in Fig.~\ref{fig:jetwidth} an error bar of width $\pm$ 2.5~$\mu$m, which is the typical distance between the two closest contours. 
We measure the width of the jet derived from the interferometric measurements
in the same way. As our estimate of the error bars for the interferometric jet width, we take roughly half the distance between the fringes ($\pm 7.5 \, \mu$m), except at 1.1~ns when the jet is still forming. 
{\blue This error estimate does not include systematic effects, such the possibility of a slight misalignment of the laser on the target \cite{Purvis_etal2010}.} 
The width of the jet in the FLASH simulations can be inferred in a precise way using the above definition  and at a large number of times.

{\blue Fig.~\ref{fig:jetwidth} shows that the FLASH and HYDRA results for this quantification of the width of the jet as a function of time agree at the earliest two times and the final time but differ somewhat at the two intermediate times. FLASH and HYDRA both over predict the width of the jet compared to the experimental measurement at 1.1 ns, but the uncertainty in the width could well be larger at this early time than our estimate. 
At the two intermediate times, the results of the HYDRA simulation lie somewhat closer to the experimental width we derived from the interferometry than do the results of the FLASH simulation, which somewhat under predict the width of the jet at 4.6~ns and after. The reason for this is unclear.
}

\subsection{Results including Radiation: Electron Temperature and Mean Ionization State}

\begin{figure}
\centerline{\includegraphics[angle=0,width=3.5in]{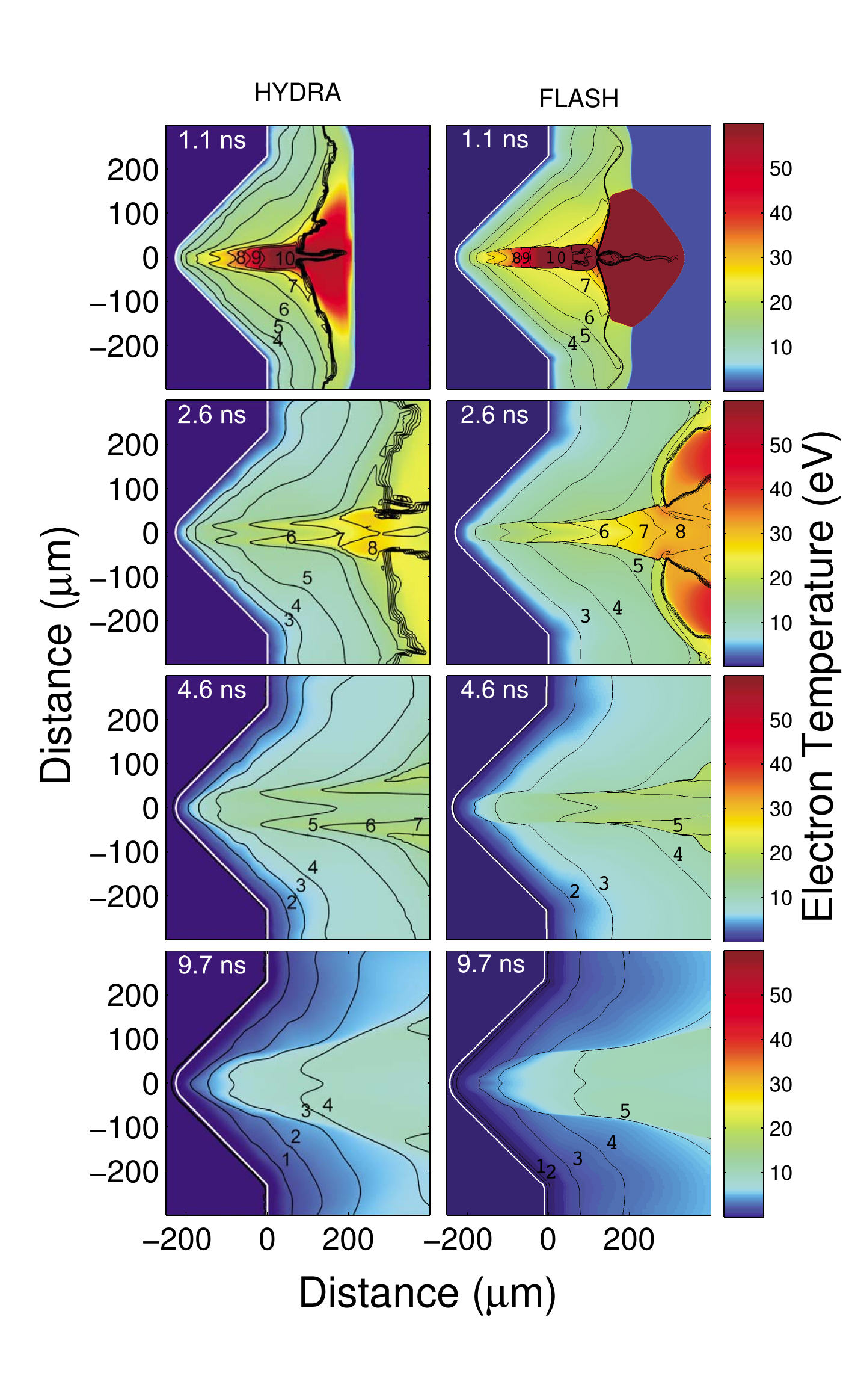}}
\vspace{-1.0cm}
\caption{
{\blue Comparison of the electron temperatures in the HYDRA (left column) and FLASH (right column) radiation-hydrodynamics simulations. 
Contours of the mean ionization state, $\bar{Z}$, are  over plotted as black lines. At 1.1~ns and 2.6~ns these contours are bunched together as closely spaced wavy, vertical lines in both the FLASH and HYDRA results. The bunching of the contours marks the position of the shock in the low density He that is produced by the outward flow of the jet and the plasma ablated from the slanted sides of the Al target.
The dark red regions to the right of these lines have a very high temperature because the composition in them is primarily very low density He.  
By 4.6 ns, the regions composed primarily of He are no longer visible in the frame. 
}
The HYDRA panels are adapted with permission from Fig.~6 in \citetalias{Grava_etal2008} with permission (copyrighted by the American Physical Society).
}\label{fig:tele}
\end{figure}

Fig.~\ref{fig:tele} compares the electron temperatures in FLASH and HYDRA at the same times reported in {\blue Fig.~\ref{fig:combined}}. 
In both the FLASH and HYDRA simulations, {\blue the interface between the expanding Al plasma and the He used to represent the vacuum} is visible at 1.1~ns and 2.6~ns. The $\bar{Z}$ contours in Fig.~\ref{fig:tele} show that at this boundary the plasma goes from a region where Al is significantly ionized to a region where $\bar{Z}$ can at most be equal to two. 
As a result, the narrow region of closely spaced contours corresponding to this transition moves steadily away from the target; by 4.6~ns the transition has left the grid. Fig.~\ref{fig:tele} indicates that at 1.1~ns the He temperatures rightward of the interface are somewhat higher in the FLASH simulation than in the HYDRA simulation. This difference could stem from a difference in how key physical processes are handled in this very low density region, including the non-equilibrium (i.e. $T_{\rm ele} \neq T_{\rm ion}$) nature of the shock. More prosaically, the He density assumed but not reported in \citetalias{Grava_etal2008} may simply be higher than what we assumed for the FLASH simulation (which was $\rho = 5 \cdot 10^{-7}$ g/cm$^{3}$).
{\blue These differences are not expected to affect the properties of the jet in the FLASH and HYDRA simulations for the reasons discussed in Section II.}

In the expanding Al plasma leftward of {\blue the} Al/He interface, the results of the FLASH and HYDRA simulations are qualitatively similar, with $T_{\rm ele}$ being slightly higher at 1.1~ns in the FLASH simulation than in the HYDRA simulation. The plasma is slightly more ionized in the FLASH simulation than in the HYDRA simulation.
The $\bar{Z}$ contours in Fig.~\ref{fig:tele} are consistent with an overall difference of $\Delta \bar{Z} \sim 1$ between the results of the two simulations. 
A closer look at the FLASH output at 1.1~ns reveals that this is true for the highest ionization state as well, and we find some regions where $\bar{Z} = 11$. Both the FLASH and HYDRA simulations agree that the mean ionization state never reaches $\bar{Z} \sim 12$, which would require much higher temperatures. \citetalias{Grava_etal2008} states {\blue that the highest mean ionization state is $\bar{Z} \sim 10$ in their HYDRA simulation and in the experiment, as shown by the absence of signatures of more-highly-ionized charge states in the time-integrated soft x-ray spectrum they obtained.  The ionization states in the HYDRA simulation therefore seem to be closer to the experiment than the FLASH simulation. Differences in the EOS models could potentially explain this result, but unfortunately we only have limited information about the assumed EOS in \citetalias{Grava_etal2008}.}

\begin{figure}
\centerline{\includegraphics[angle=0,width=3.5in]{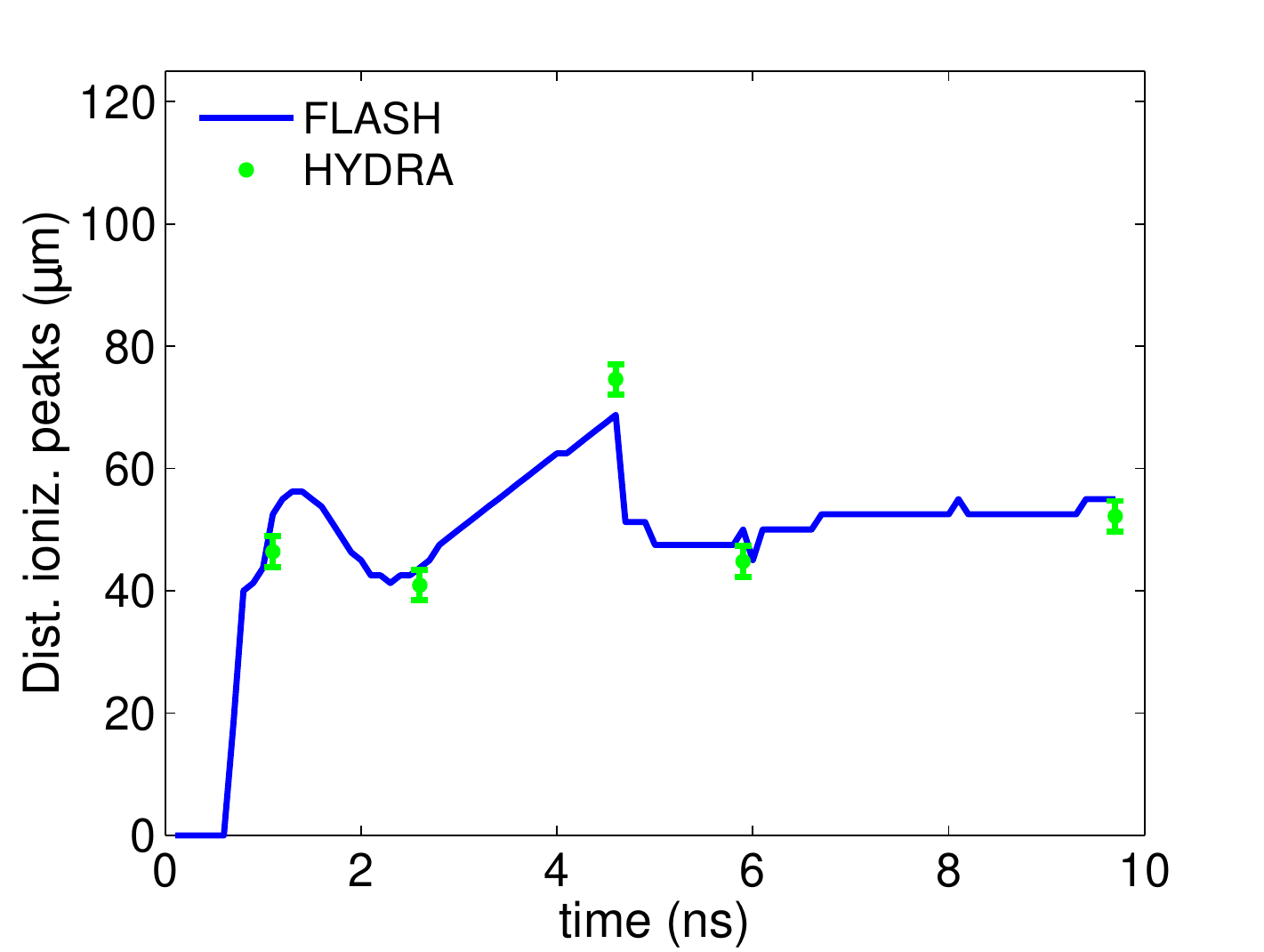}}
\vspace{-0.25cm}
\caption{A quantitative comparison of the distance between the peaks in 
the mean ionization state ($\bar{Z}$) on a line perpendicular to the laser axis that is centered on (0,0). 
The HYDRA measurements (green points) come from the $\bar{Z}$ contours in Fig.~\ref{fig:tele}, which are derived 
from \citetalias{Grava_etal2008}. Measurements from the FLASH simulation (solid blue line) are presented
at finely spaced intervals in time.}\label{fig:jetwidth_zbar}
\end{figure}

The contours of $\bar{Z}$ can be used as another measure of the width of the jet versus time to quantify the level of agreement in Fig.~\ref{fig:tele}. Aluminum ablating from the walls expands and collides with plasma flowing outward on axis, creating peaks in $T_{\rm ele}$ and $\bar{Z}$ just above and below the axis of the laser beam instead of on axis. This produces the double-horned features in $T_{\rm ele}$ and $\bar{Z}$ seen  in Fig.~\ref{fig:tele} prior to 5~ns. The distance between the peaks in $\bar{Z}$ can be measured on a line perpendicular to the laser axis centered at (0,0) for the FLASH and HYDRA simulations, giving a different determination of the width of the jet. Fig.~\ref{fig:jetwidth_zbar} compares this measurement for the HYDRA simulation to a very precise measurement for the FLASH simulation at many times. Before 4.6~ns, this definition of the jet width gives results similar to those in Fig.~\ref{fig:jetwidth}. 
However, after 5~ns the double-horned feature in $\bar{Z}$ is much less pronounced because the Al plasma arriving along the axis of the laser beam is no longer plasma {\blue at the center of the groove in the Al target} that was directly heated by the
laser (c.f. Fig.~10 in \citetalias{Grava_etal2008}). 
The speed and momenta of the plasma that is ablating from the walls and arriving along the axis of the laser beam has decreased significantly, causing a transition in which the double-horned feature nearly disappears and the width of the jet suddenly decreases. The abrupt decrease in the width of the jet seen in Fig.~\ref{fig:jetwidth_zbar} at around 5~ns is due to this transition.

The results of the FLASH and HYDRA simulations agree well on the width of the jet as measured by the distance between the two peaks in $\bar{Z}$, despite the overall difference of $\Delta \bar{Z} \sim 1$ in the mean ionization state discussed earlier. 
Since no experimental measurement of the mean ionization state of the plasma is available in \citetalias{Grava_etal2008}, these results constitute a code-to-code comparison between FLASH and HYDRA only.

\subsection{Results including Radiation: Total Pressure}

\begin{figure}
\centerline{\includegraphics[angle=0,width=3.5in]{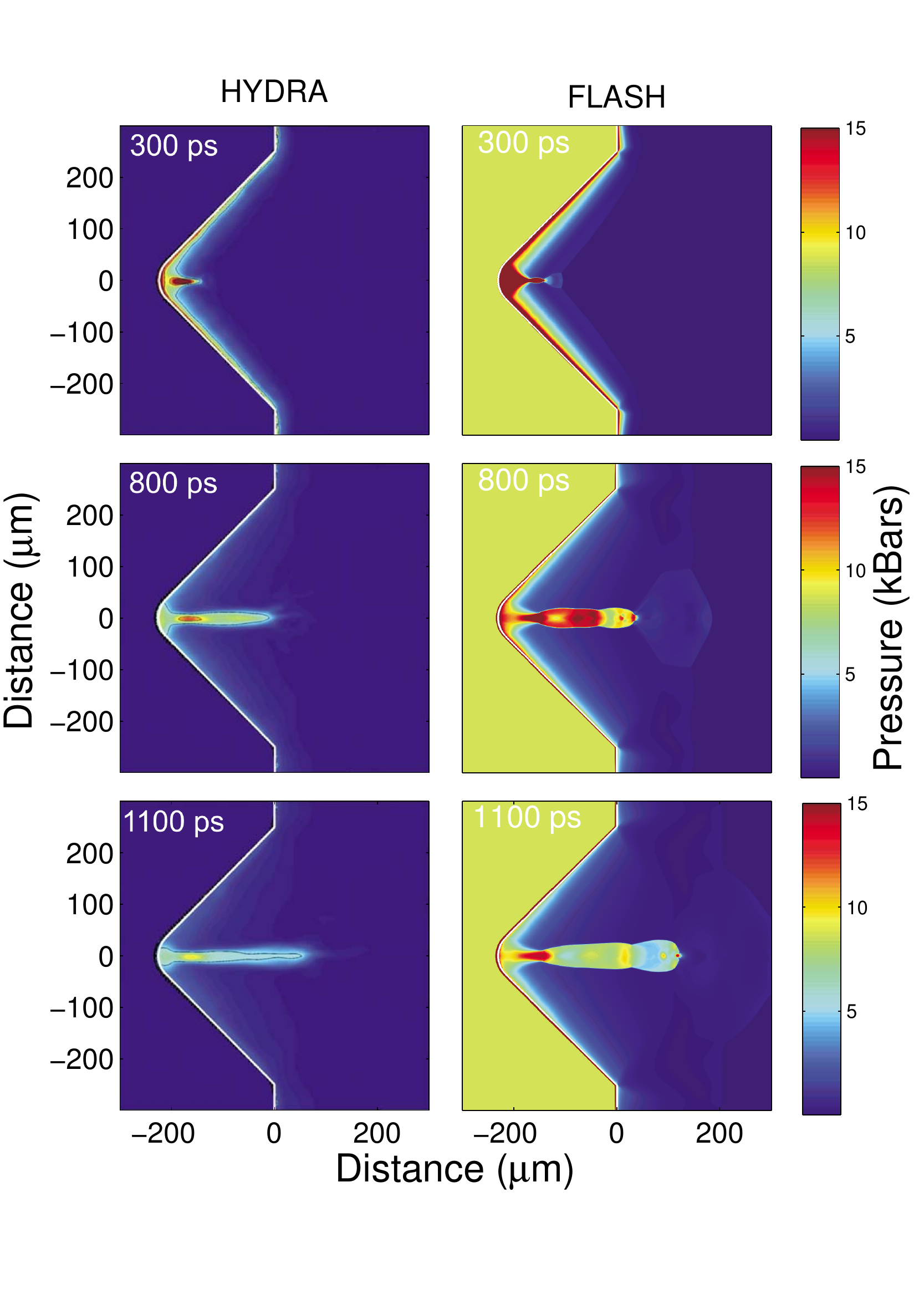}}
\vspace{-1cm}
\caption{
Comparing total plasma pressure at early times from radiation-hydrodynamics simulations
from HYDRA (left column) and FLASH (right column). The FLASH simulation was performed using PROPACEOS opacity and EOS data. HYDRA panels are adapted from Fig.~8 in \citetalias{Grava_etal2008} with permission (copyrighted by the American Physical Society).
}\label{fig:pres}
\end{figure}

Fig.~\ref{fig:pres}  compares the total plasma pressure at early times in the FLASH and HYDRA simulations with radiation. While FLASH and HYDRA use EOS models that are related to or are the same as the QEOS model constructed by More et al. \cite{QEOS}, 
{\blue
there are differences in the implementation. Specifically, 
in HYDRA the total pressure at solid density, i.e. in the initial Al target}, is vanishingly small, as one expects for a cold solid.
FLASH uses the (QEOS-based) PROPACEOS EOS which achieves vanishingly small total pressures for cold solids, in part, by allowing the electron pressure to be negative. FLASH currently requires positive electron and ion pressures in solving the momentum equation for the fluid and in interpolating EOS data. We therefore set the few negative electron pressures that exist in the PROPACEOS table to a very small, positive value. Thus the $\sim 7$~kBar (olive-colored) pressures well into the target in the FLASH results reflect only the {\it ion} pressure reported from PROPACEOS.
Bearing in mind, from a hydrodynamic point of view, only pressure {\it differences} matter and that the laser-heated Al quickly enters a regime where the total pressure is well above zero, it is understandable that this difference seems not to have significantly affected the agreement between the two codes.

{\blue
\subsection{Conclusion of the Code-to-Code Comparison and Validation Effort}
}
{\blue
The results of the FLASH and HYDRA simulations of the \citetalias{Grava_etal2008} experiment are in reasonable agreement, but differences do exist.  The differences appear to be explainable by somewhat higher temperatures near the surface of the Al target, especially at the center of the groove, in the FLASH simulations relative to the HYDRA simulations.  These somewhat higher temperatures lead to higher degrees of ionization and therefore larger electron number densities and higher pressures close to the surface of the target in the FLASH simulations. 

The larger electron number densities present in the FLASH simulations compared to the HYDRA simulations are evident in the ``no radiation" cases shown in the two left-most columns of Fig.~\ref{fig:combined}.  
The jet in the FLASH simulation is both wider and moves more rapidly away from the target compared to the HYDRA results. This is due to an overall higher pressure in FLASH which can be seen in Fig.~\ref{fig:pres}.


The comparisons of the results for FLASH and HYDRA simulations with radiation diffusion shown in the panels in the two right-most columns of Fig.~\ref{fig:combined} are also consistent with this picture.  The jet in the FLASH simulation is again wider and moves more rapidly away from the target through 2.6 ns due to the higher pressure.  After this, the jet in the FLASH simulations is narrower, which we attribute to greater radiative cooling in the FLASH simulation as a result of the higher degree of ionization.

The somewhat higher electron temperatures in the FLASH simulation with radiation diffusion compared to the HYDRA simulation with radiation is evident in the fact that at  early times ($\lesssim$ 500 ps) the majority of the jet in the FLASH simulation has $\bar{Z} \sim 11$ (not shown), whereas according to \citetalias{Grava_etal2008} the maximum ion state in the HYDRA simulations is $\bar{Z} = 10$ and extreme UV spectroscopy data presented in \citetalias{Grava_etal2008} strongly indicates that higher ionization states were not present. By 2.6~ns, which is well after the laser has turned off, the ionization states in the FLASH simulation have decreased to values that are no longer in disagreement with the spectroscopic data. 
The higher pressures due to the higher electron temperatures and degree of ionization at the surface of the Al target at early times is evident in Fig.~\ref{fig:pres}.

The higher temperatures at the surface of the Al target in the FLASH simulations could be the principal source of the differences between the results of the FLASH simulations and the HYDRA simulations.  However, the treatment of the laser beam and the rate  at which the laser beam deposits energy should be identical in FLASH and HYDRA, and therefore are not likely to be the source of the difference.  The treatment of the pressure in the cold, solid Al target is slightly different (as highlighted in Fig.~\ref{fig:pres}), but as discussed in Sec.~\ref{sec:comparison-flash-hydra}, it is unlikely this small difference can account for the higher electron temperature at and above the surface of the Al.

Another possible explanation of the higher temperatures at the surface of the Al target in the FLASH simulations compared to the HYDRA simulations is the markedly different treatments of the computational mesh in the two codes.  The critical density at which the energy in the laser is absorbed in the Al target is highly refined in the FLASH simulations as a result of its use of AMR.  The rapidly expanding flow at the surface of the Al target causes the mesh to expand rapidly in this region in codes that use the ALE approach as does HYDRA.  This presents a challenge in terms of maintaining sufficient spatial resolution in this important region, a challenge that can be overcome by frequent rezoning.  Insufficient resolution can result in less laser energy deposited in the target.  However, the better agreement between the results of the HYDRA simulation with the experimental results from \citetalias{Grava_etal2008} compared to the results of the FLASH simulation strongly suggests this is not the case.


The fact that opacity models typically disagree at the level of a few to tens of percent, as discussed in Sec.~\ref{sec:comparison-flash-hydra}, could potentially explain some differences between FLASH and HYDRA. If opacities were causing a difference between the codes one would expect that the radiative results would be very different for FLASH and HYDRA even if the non-radiative results are similar.  But, overall, the non-radiative results for the electron density from FLASH and HYDRA agree about as well as the radiative results (Fig.~\ref{fig:combined}). 

The remaining possibility is that the observed differences between the FLASH and HYDRA simulations is due to differences in the EOS, even though the HYDRA simulations of the \citetalias{Grava_etal2008} experiment use the QEOS model and the PROPACEOS EOS used in the FLASH simulations also use a QEOS-based model.

We conclude that future, more detailed code-to-code comparisons that have access to greater knowledge of the components of the HYDRA code, and so can determine more clearly the effects of the various components in the FLASH and HYDRA codes, will be necessary to identify with confidence the source or sources of the differences between the FLASH and HYDRA simulations of the \citetalias{Grava_etal2008} experiment discussed in this paper.
}

\begin{figure*}[ht!]
\includegraphics[width=7in]{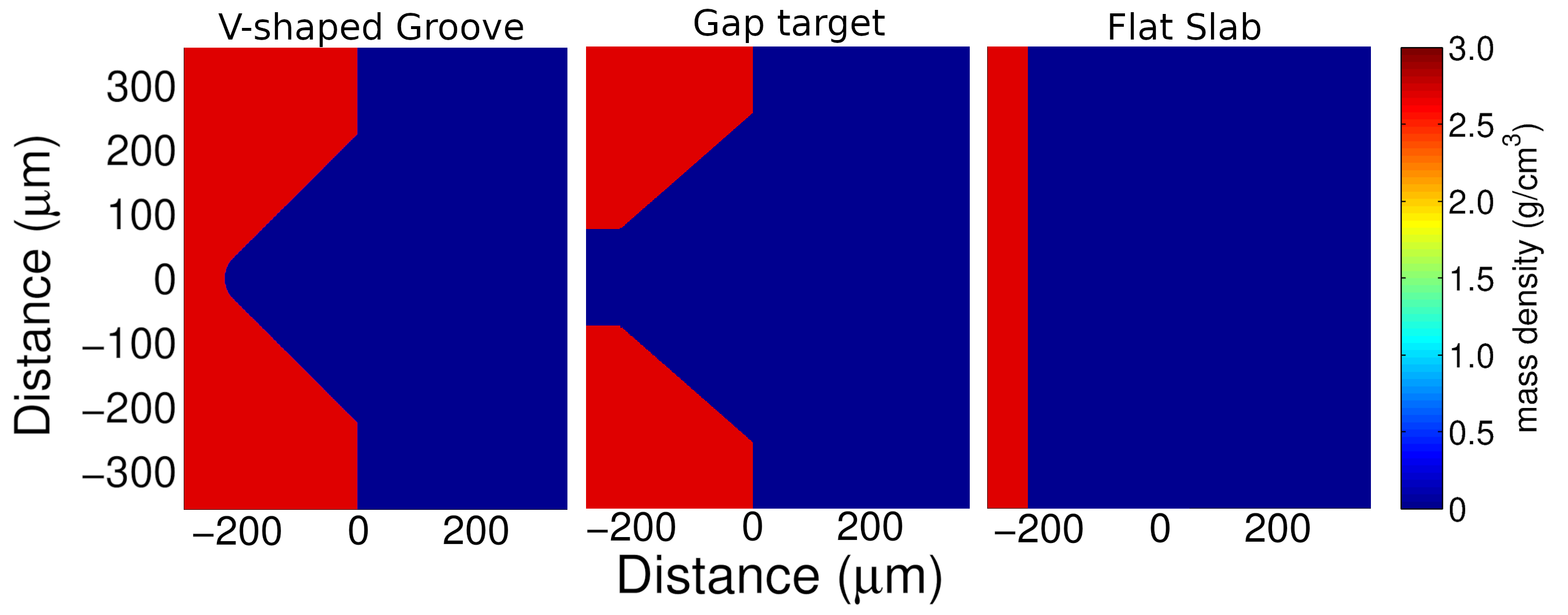}
\vspace{-0.5cm}
\caption{{\blue  The three different targets that we consider in Sec~\ref{sec:jet}. The V-shaped Groove target (left panel) was investigated in earlier parts of this paper, following work by \citetalias{Grava_etal2008}. The Gap target design (center panel) was investigated in NOVA experiments by \citetalias{Wan_etal1997}. The flat slab target (right panel) is for comparison.} }\label{fig:targets}
\end{figure*}

\section{Formation and Properties of the Jet in the GRAVA Experiment}
\label{sec:jet}


The \citetalias{Grava_etal2008} experiment had two objectives:  (1) to obtain data that would make possible an important additional validation test of radiative hydrodynamics codes in the wake of the apparent failure of LASNEX \cite{LASNEX} simulations to reproduce a similar experiment done using the Nova laser \citetalias{Wan_etal1997}; and (2) to create a jet analogous to astrophysics jets, following Stone et al. \cite{Stone_etal2000}.  
Having validated FLASH for the \citetalias{Grava_etal2008} experiment, we now use FLASH simulations to better understand the formation and properties of the jet.  
We focus on early times ($\leq$~1.1~ns) and to inform our discussion we use comparisons between (1) the FLASH simulations of this experiment presented earlier in  Sec.~\ref{sec:grava}, {\blue (2) a FLASH simulation using the same laser beam with a V-shaped groove but with the inner $\pm 75 \mu$m section of the target removed, and (3) a FLASH simulation involving a flat Al target irradiated by the same rectangular laser beam used in the \citetalias{Grava_etal2008} experiment. These three targets are illustrated in Fig.~\ref{fig:targets}.} {\blue The second} configuration resembles an earlier experiment done by  \citetalias{Wan_etal1997} in which two slabs with a gap between them were oriented perpendicular to each other and irradiated by beams of the NOVA laser. \citetalias{Grava_etal2008} cites this experiment as an important motivation for their work.

A number of questions arise in drawing parallels between the self-colliding, ablating plasma in \citetalias{Grava_etal2008} and astrophysical jets.  While in both the astrophysical and laboratory context the radiative cooling timescale may be similar in magnitude to the hydrodynamic expansion timescale (and therefore the adiabatic cooling timescale), how similar are these situations in other respects?  Here we address three specific questions about the formation and properties of the jets seen in the laser experiments whose answers enable us to compare them with astrophysical jets:

\begin{enumerate}

\item
What determines the physical conditions (e.g., the density, temperature, and velocity) in the core of the jet?

\item
Is the collimating effect of the plasma ablating from the angular sides of the groove due to thermal pressure, i.e., the internal energy of the ablating plasma, or ram pressure, i.e., the component of the momentum of the ablating plasma perpendicular to the mid-plane of the experiment?

\item
Does the collimation of the flow by the plasma ablating from the angular sides of the groove and the entrainment of this plasma in the resulting jet increase or decrease the velocity of the jet?

\end{enumerate}

Laboratory astrophysical jet experiments remain an active field of inquiry (e.g. \cite{Gao_etal2019}) and we hope that addressing these questions will aid in future research.

\subsection{What determines the physical conditions in the core of the jet?}

Figures \ref{fig:jetwidth} and \ref{fig:jetwidth_zbar} show that the width of the jet at early times ($\leq$~1.1~ns) is similar to the width of the rounded region at the center of the V-shaped groove which is relatively flat and relatively perpendicular to the laser beam illuminating the target.  The incident laser beam has a Gaussian cross section with a FWHM of 360~$\mu$m which implies that most of the laser energy is deposited within $\pm 100~\mu$m of the laser axis. Accordingly, the intensity of the laser beam will be much greater on the relatively flat portion in the middle of the V-shaped groove than on the sides of the groove. In this way, the interaction near the center of the groove resembles the simple case of an Al slab (i.e. flat) target. If the physical mechanism producing the jet is the same in for both the V-shaped grove and slab target then one would expect that the velocity at various locations along the core of the jet should be similar in the two cases.

\begin{figure*}
\centerline{\includegraphics[angle=0,width=7.1in]{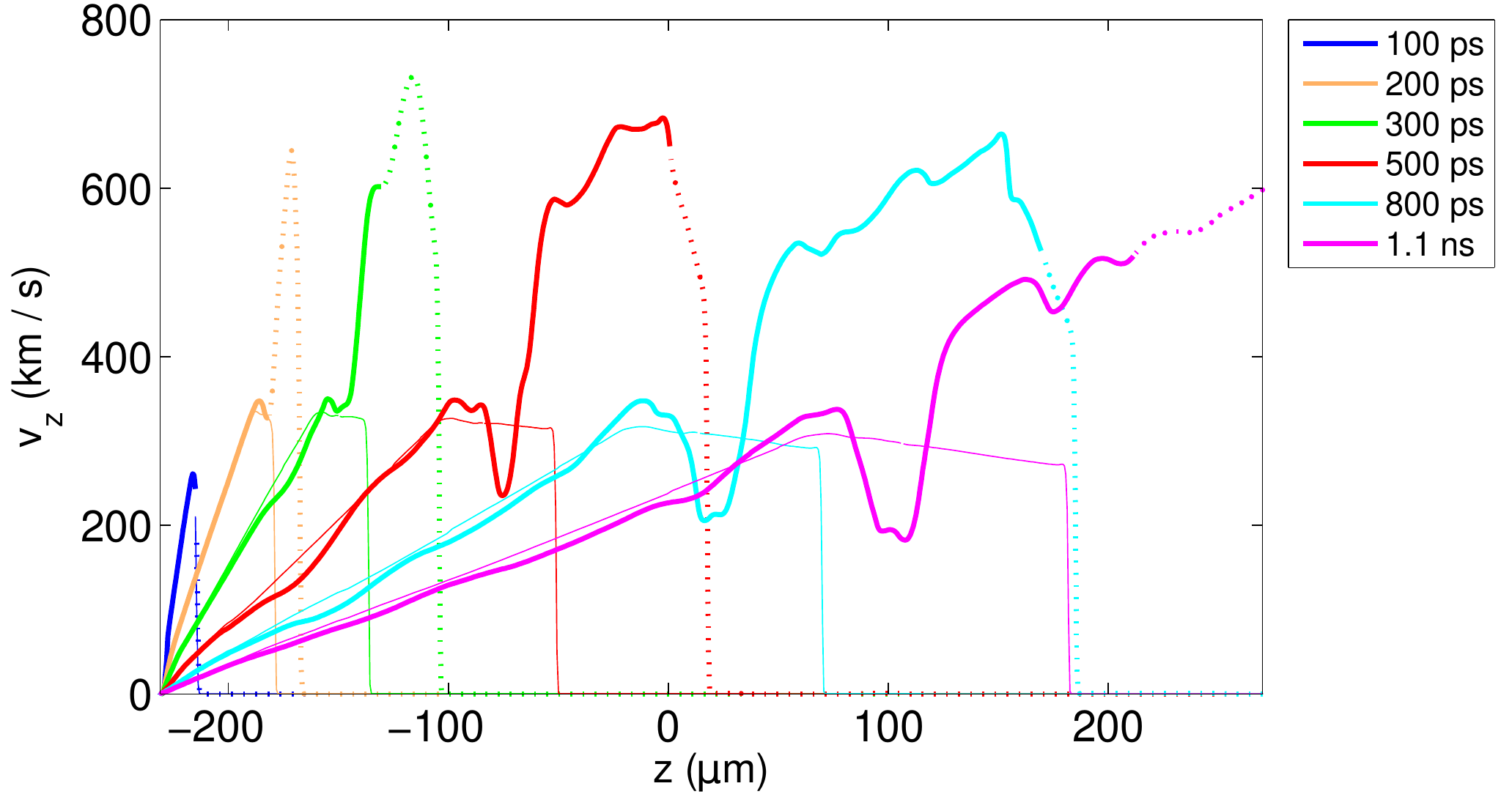}}
\vspace{-0.3cm}
\caption{
Lineouts along the laser axis of the component of the velocity parallel to the laser axis at various times. The thick solid lines show results from the FLASH simulation of a V-shaped groove target described in  Sec. \S~\ref{sec:grava}. These thick solid lines become dotted lines at the point where the cell material is mostly very low density He instead of Al. Thin solid lines (which become dashed lines at the Al/He transition) show the same measurements from a simulation where a flat target of the same material is irradiated with the \citetalias{Grava_etal2008} laser pulse.
}\label{fig:vely1}
\end{figure*}

To investigate this possibility, we compared lineouts of the velocity along the mid-plane for the two problems at five relatively early times; i.e., $\leq$ 1.1 ns. These are shown in  Fig. \ref{fig:vely1}.  Ignoring the very high ($v_z~\gtrsim~350$~km/s) velocities of very low density gas and focusing on $v_z \lesssim 350$~km/s, {\it the velocity profiles for the two problems agree closely at all five times, supporting our conjecture that the mechanism producing them is the same}. The differences seen between the two simulations at very high velocities ($v_z \gtrsim 350$~km/s) are at very low densities and near or approaching the transition from cells that are mostly Al to cells that are mostly very low-density He. This transition is marked by a change in line type from thick solid to dotted lines for the V-shaped target or from thin solid to dashed lines for the flat target. Because the He serves only as an approximation to vacuum conditions, the results  near this very low-density interface are not relevant to the questions we are concerned with in this section.

\begin{figure*}
\centerline{\includegraphics[angle=0,width=7.1in]{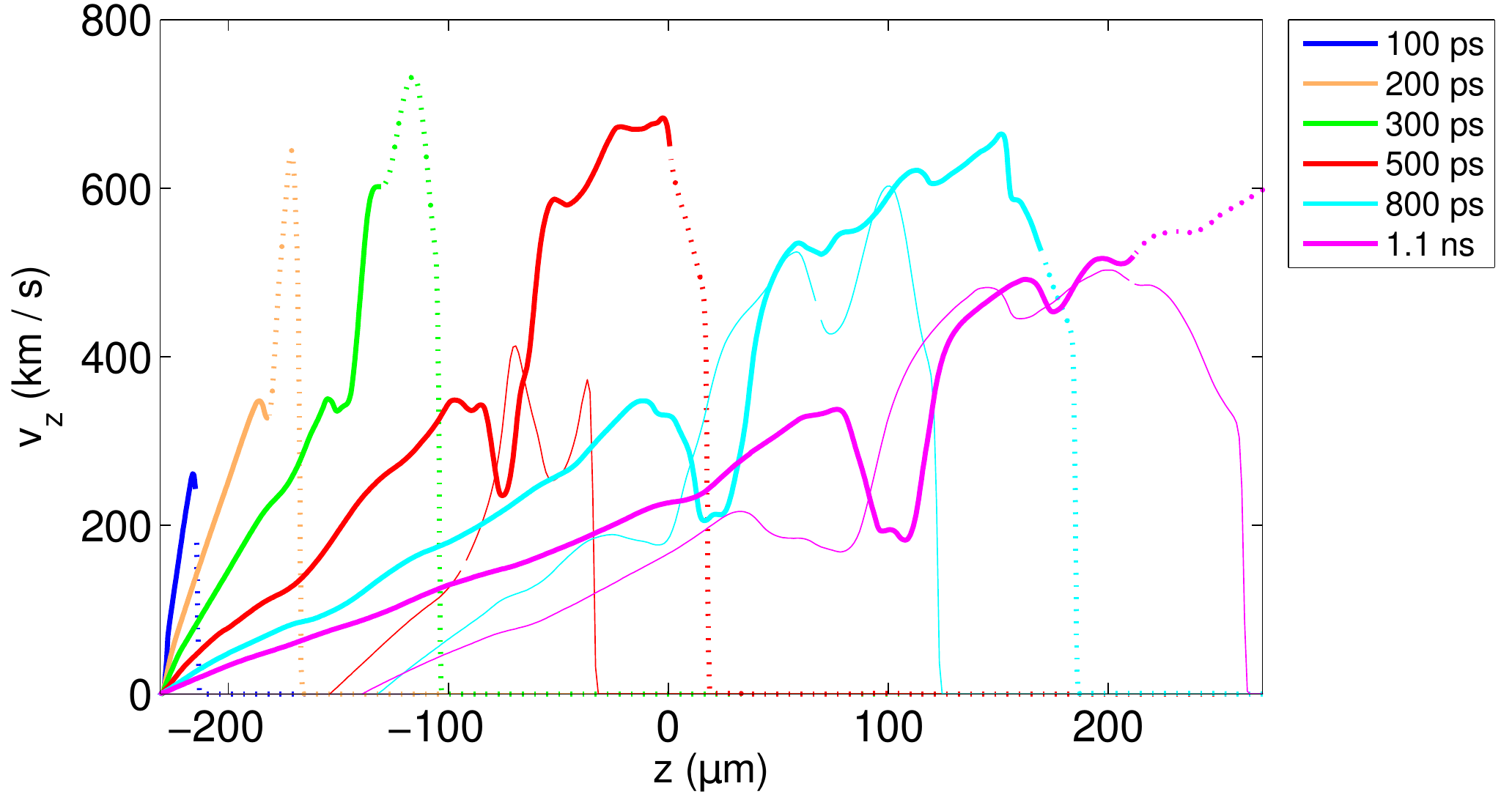}}
\vspace{-0.3cm}
\caption{
Lineouts along the laser axis of the component of the velocity parallel to the laser axis at various times. The thick solid lines show measurements from the FLASH simulation of a V-shaped groove target described in Sec.~\ref{sec:grava}. These thick solid lines become dotted lines at the Al/He transition. Thin solid lines (which likewise become dashed lines at the Al/He transition) show the same measurements from a simulation where, similar to the target geometry of \citetalias{Wan_etal1997}, a large gap of material is missing from the center of the V-shaped groove. This target is irradiated by the \citetalias{Grava_etal2008} laser pulse. 
}\label{fig:vely2}
\end{figure*}

We also compared the lineouts of the velocity along the mid-plane for the Al target with a V-shaped groove to a target consisting of two Al slabs oriented perpendicular to each other with a gap between them, similar to the \citetalias{Wan_etal1997} experiment. This comparison allows us to contrast the properties of the jet produced by a V-shaped groove that has a relatively flat portion near the mid-plane and one that does not.  Figure \ref{fig:vely2} shows the results  of the simulation for the V-shaped groove target and that for the \citetalias{Wan_etal1997}-like target. As in Fig.~\ref{fig:vely1} the change from a solid line to a dotted or dashed line indicates the transition from mostly Al to mostly He. 
Focusing on the thin and thick solid lines that correspond to Al target material, the two velocity profiles differ greatly.  The absence of a relatively flat portion of the target near the mid-plane means that formation of the jet is delayed until the plasma ablating from the sloping sides of the target has had time to meet at the mid-plane.  Furthermore, the velocity profile of the jet is much shallower and its maximum velocity is much smaller.  These results provide further support for the hypothesis that the physical mechanism producing the jet in \citetalias{Grava_etal2008} is the same as in the slab problem.

\begin{figure*}
\centerline{\includegraphics[angle=0,width=7.0in]{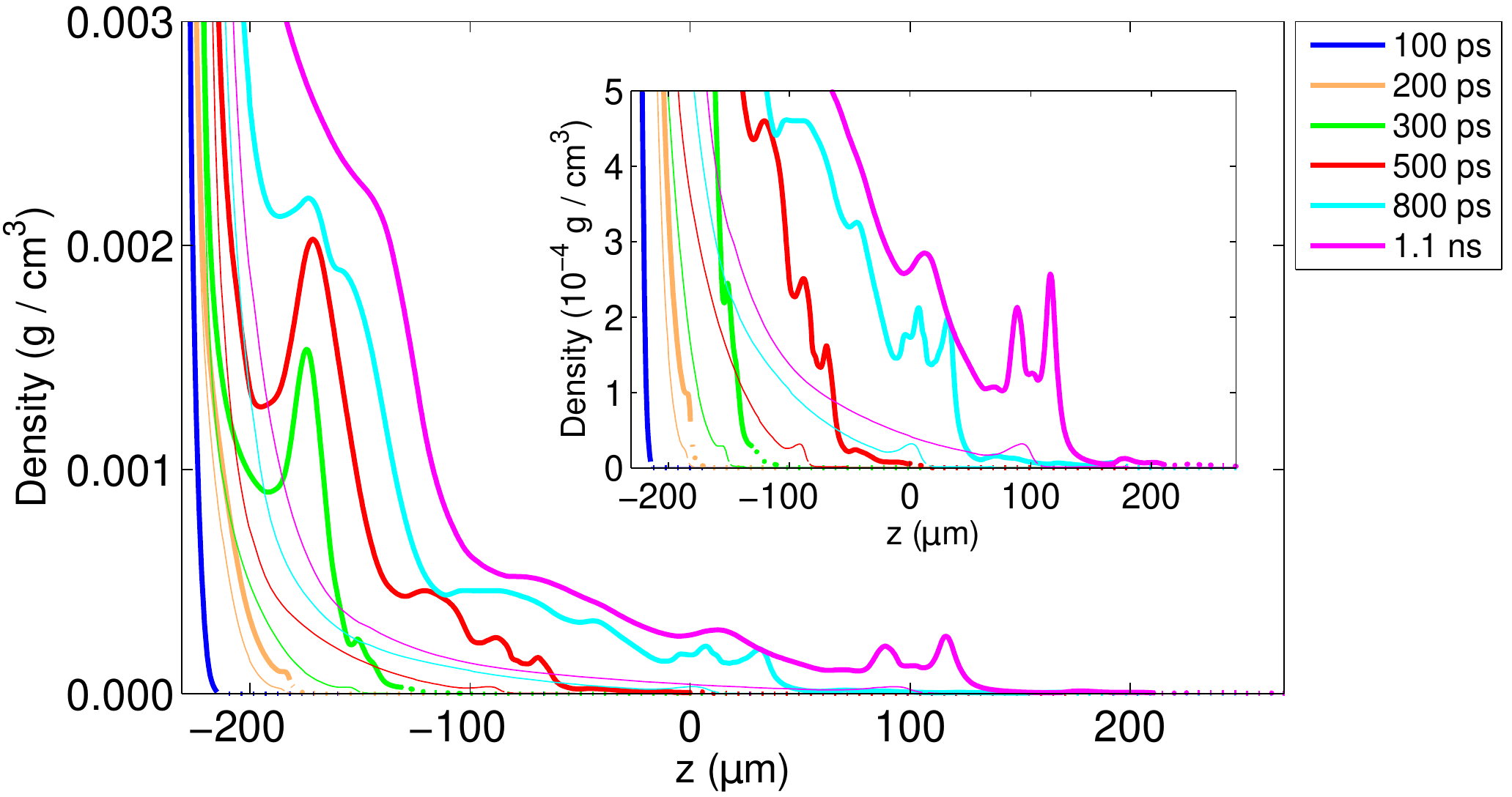}}
\vspace{-0.3cm}
\caption{
Lineouts of the mass density along the laser axis at various times. The thick solid lines show measurements from the FLASH simulation of a V-shaped groove target described in Sec.~\ref{sec:grava}. Thin solid lines show the same measurements from a simulation where a flat target of the same material is irradiated with the \citetalias{Grava_etal2008} laser pulse. As in Fig.~\ref{fig:vely1}, the change from a solid line to a dotted or dashed line indicates the Al/He transition. {\blue An inset figure highlights densities below $5\cdot 10^{-4}$~g/cm$^3$ in these simulations using the same line type scheme as the larger figure.}
}\label{fig:dens1}
\end{figure*}

\begin{figure*}
\centerline{\includegraphics[angle=0,width=7.0in]{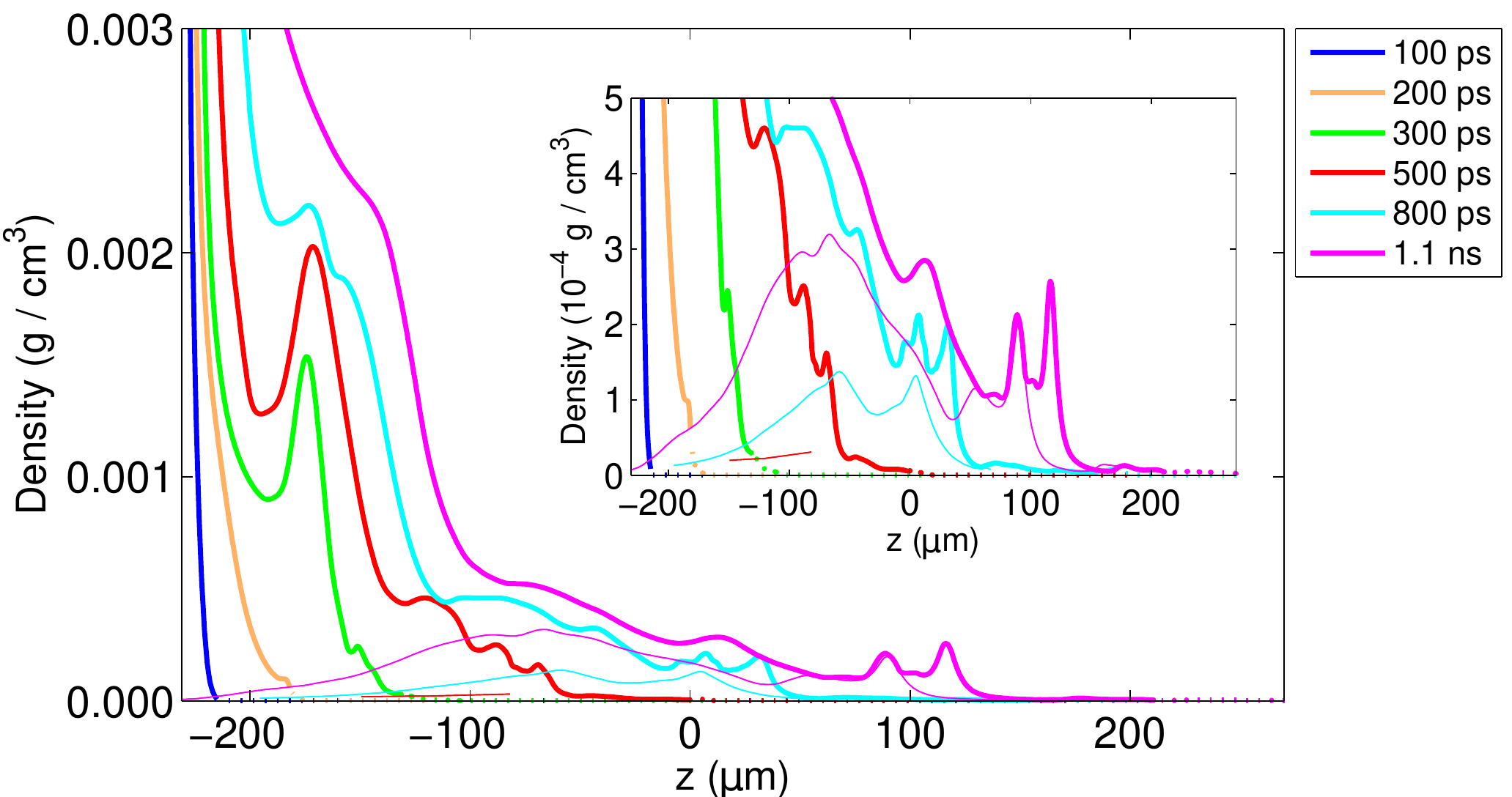}}
\vspace{-0.3cm}
\caption{
Lineouts of the mass density along the laser axis at various times. The thick solid lines show measurements from the FLASH simulations of a V-shaped groove target described in Sec.~\ref{sec:grava}. Dashed lines show the same measurements from a simulation where, similar to the target geometry of \citetalias{Wan_etal1997}, a large gap of material is missing from the center of the V-shaped groove. This target is irradiated by the \citetalias{Grava_etal2008} laser pulse. As in Fig.~\ref{fig:vely2}, the change from a solid line to a dotted or dashed line indicates the Al/He transition. {\blue An inset figure highlights densities below $5\cdot 10^{-4}$~g/cm$^3$ in these simulations using the same line type scheme as the larger figure.}
}\label{fig:dens2}
\end{figure*}

Figure \ref{fig:dens1} compares lineouts of the density along the mid-plane for the V-shaped groove target (thick solid lines) and a flat Al slab target (thin solid lines), while Fig.~\ref{fig:dens2} compares the same measurement from the V-shaped groove target simulation (thick solid lines) to results from a target consisting of two Al slabs oriented perpendicular to each other with a gap between them (thin solid lines), similar to the \citetalias{Wan_etal1997} experiment. As in Figs.~\ref{fig:vely1} and \ref{fig:vely2} these lines become dotted or dashed when the cells are mostly He instead of Al.

Clearly, ablation from the sloping sides of the V-shaped groove confines the flow, as discussed by  \citetalias{Wan_etal1997} and \citetalias{Grava_etal2008}, greatly increasing the density in the jet relative to the density in the case of the flat Al slab target, where the flow can expand laterally as well as away from the surface of the target.  Collimation of the flow by the plasma ablating from the sloping sides of the V-shaped groove in \citetalias{Grava_etal2008} raises the question of whether the collimation is due primarily to thermal pressure or to ram pressure.  We now address this question.

\subsection{Is the collimating effect of the plasma ablating from the angular sides of the groove due to thermal pressure or ram pressure?}



\begin{figure*}
    \centering
    \includegraphics[width=7in]{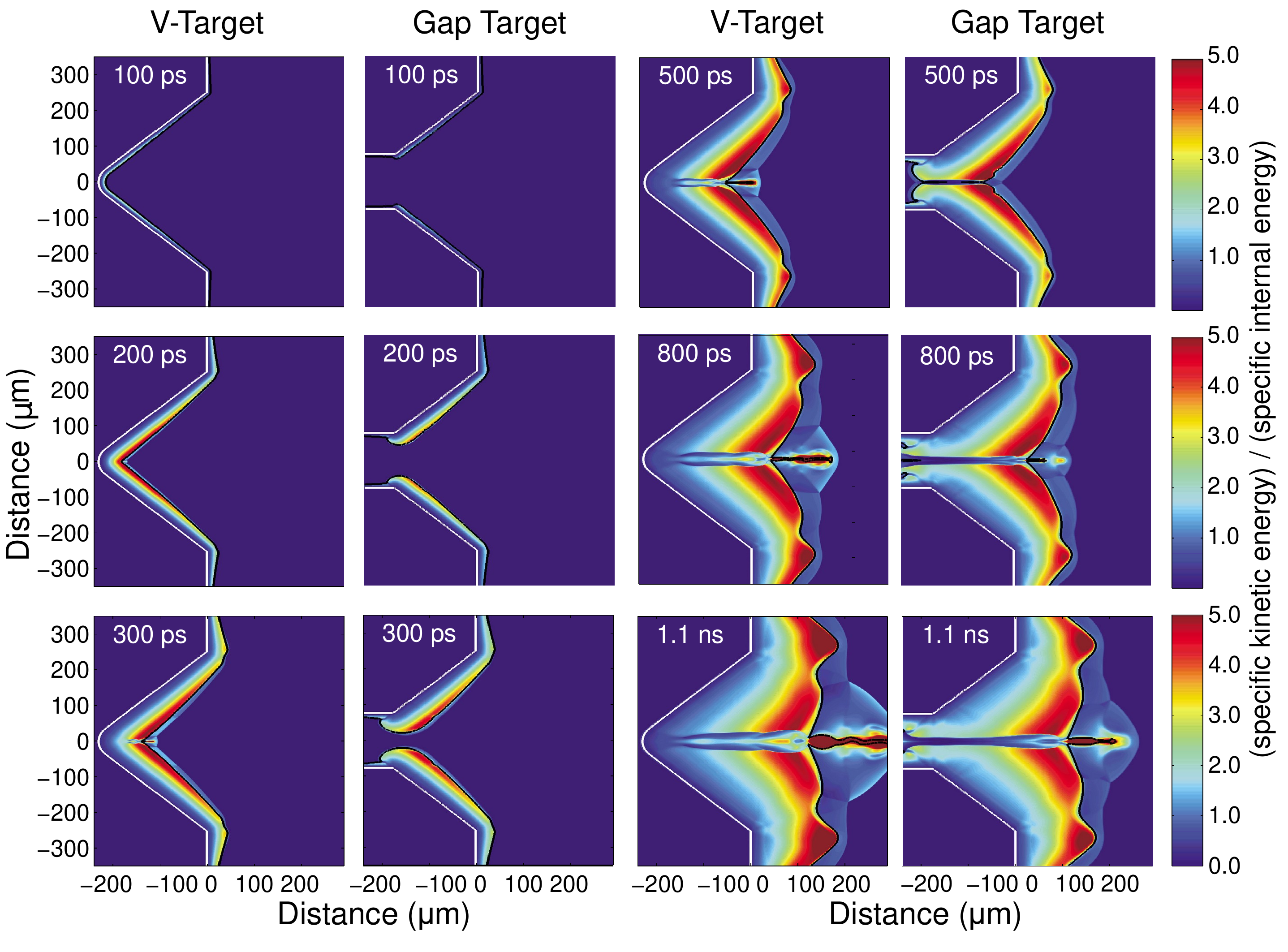}
    \vspace{-0.5cm}
    \caption{{\blue Plots (labeled "V-Target") showing the ratio of $e_{\rm kin} / e_{\rm int}$ at various times for the FLASH simulation of the V-shaped groove target described in Sec.~\ref{sec:grava} and plots (labeled "Gap Target") showing the same ratio for the FLASH simulation of a V-shaped target with a gap in the center (similar to \citetalias{Wan_etal1997}).  Both targets are irradiated by the \citetalias{Grava_etal2008} laser pulse.  In each panel, the original target location is shown with a white line and the transition from cells that are mostly Al to cells that are mostly He is indicated with a solid black line.}
    }
    \label{fig:ekin_combined}
\end{figure*}

\begin{figure}
\centerline{\includegraphics[angle=0,width=3.5in]{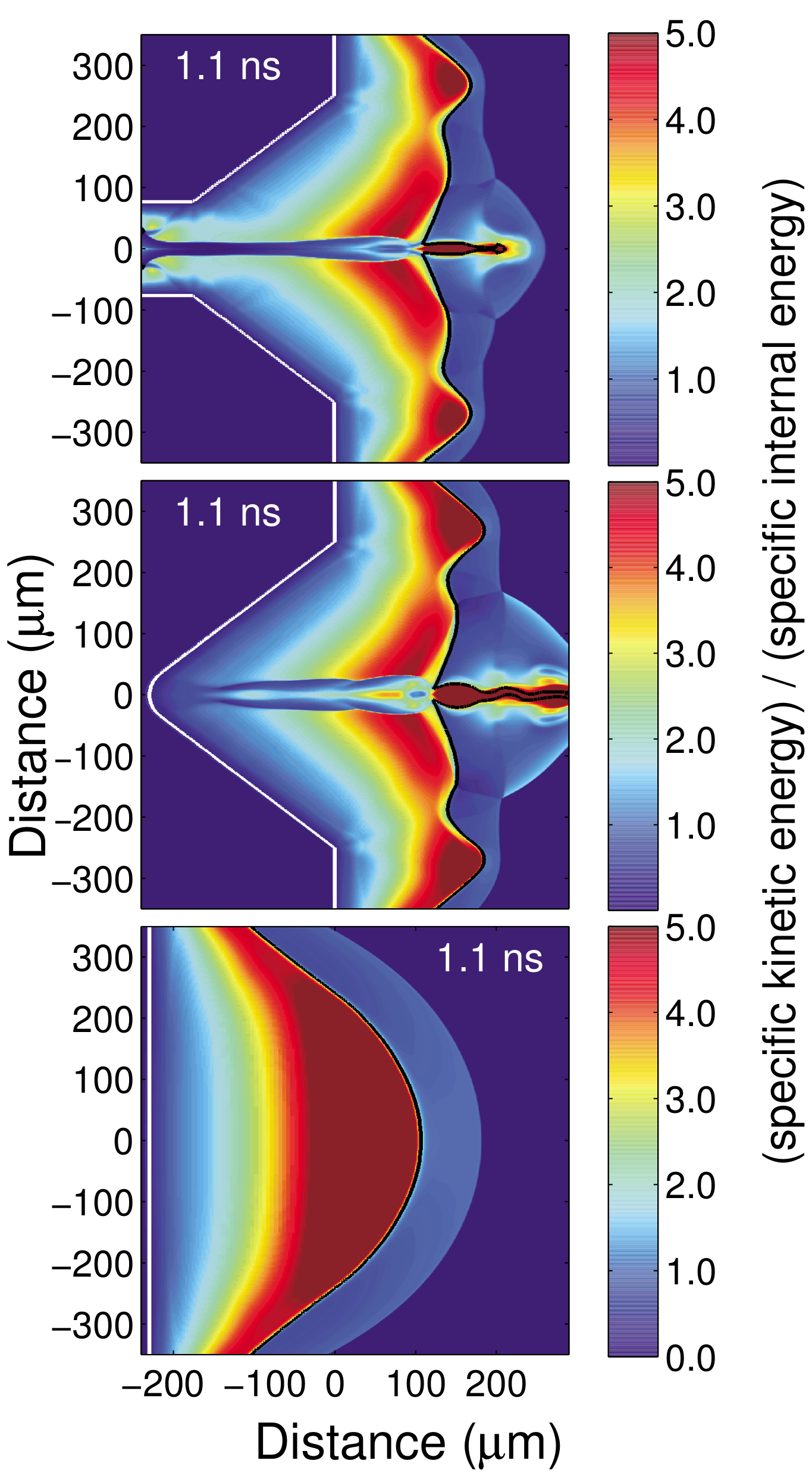}}
\vspace{-0.25cm}
\caption{
Plots showing the ratio of $e_{\rm kin} / e_{\rm int}$ at 1.1~ns for three different FLASH simulations: (Top) the V-shaped groove simulation described in Sec.~\ref{sec:grava}, (middle) results from a V-shaped groove with a gap as in \citetalias{Wan_etal1997} (bottom) results from a FLASH simulation of a flat target that is likewise irradiated by the \citetalias{Grava_etal2008} laser pulse. In each plot the original target location is shown with a thick white line. The transition from mostly Al to mostly He cells is indicated with a solid black line in each panel.
}\label{fig:ekin_nomid_nc_flat}
\end{figure}


To address this question, we calculate the ratio of the specific kinetic energy,
\begin{equation}
e_{\rm kin} = \frac{1}{2} |\vec{v}|^2
\end{equation}
to the total specific internal energy,
\begin{equation}
e_{\rm int} = e_{\rm ele} + e_{\rm ion}.
\end{equation}
If $e_{\rm kin} / e_{\rm int} \gg 1$, the kinetic energy of the ablation flow is dominant. On the other hand, if $e_{\rm kin} / e_{\rm int} \lesssim 1$, the internal energy due to the temperature of the plasma is dominant.

Figure~\ref{fig:ekin_combined} shows the ratio $e_{\rm kin}/e_{\rm int}$ throughout the computational domain at six different times for the V-shaped groove {\blue and gap targets}. 
Figure~\ref{fig:ekin_nomid_nc_flat} compares the these measurements at 1.1~ns to measurements from the simulation of a flat Al slab target. 
In Figs.~\ref{fig:ekin_combined} - \ref{fig:ekin_nomid_nc_flat}, we see a similar behavior:  even at very early times, in the plasma very near the target, the internal energy of the plasma dominates the kinetic energy of the bulk flow away from the target, but further out the kinetic energy of the bulk flow dominates the internal energy.  Once the laser turns off, the region where the kinetic energy of the bulk flow dominates the gas internal energy begins to grow substantially larger.

Examining the properties of the ablation flow as it approaches the mid-plane, we see that the internal energy of the plasma dominates at distances $< ~50~\mu$m from the target, but the kinetic energy of the bulk flow dominates at all larger distances.  This indicates that the collimation of the jet is due primarily to ram pressure except very near the target, where it is due primarily to thermal pressure of the hot plasma.  

The plasma in the jet is heated when the plasma ablating from the sloping sides of the target in both the \citetalias{Grava_etal2008} and the \citetalias{Wan_etal1997}-like experiments collides in the center, converting kinetic energy to thermal energy. This process is simplest in the \cite{Wan_etal1997} target and this collision produces a low ratio of $e_{\rm kin}/e_{\rm int}$ along the center. 

The {\blue V-shaped Groove target results} shown in Fig.~\ref{fig:ekin_combined} {\blue have} a more complex structure to $e_{\rm kin}/e_{\rm int}$ than the other cases. Material from the target collides in the center, but there is also material originating from the middle of the V-shaped groove that is moving rapidly away from the target. 
The result is a complex lateral structure within the jet in which $e_{\rm kin}/e_{\rm int}$ is large in the core of the jet and small at its edges, and large again in the ablating plasma above and below the jet.  This is the origin of the double-horn structure evident at very late times in the electron density, which can be seen in Figs. \ref{fig:combined}-\ref{fig:jetwidth}, and in the ionization state, which can be seen in Figs. \ref{fig:tele}-\ref{fig:jetwidth_zbar}.


{\blue Although the lateral structure of the plasma flow in the \citetalias{Grava_etal2008} experiment is complex, the jet that is produced in the experiment is very narrow, the ratio $e_{\rm kin}/e_{\rm int}$ is large along the axis of the jet, and this ratio decreases laterally away from the axis, all of which are properties of astrophysical jets \cite[e.g.][]{Stone_Norman1993}. 
In contrast, the gap target results produce only a modest ratio of $e_{\rm kin}/e_{\rm int}$ along the axis of the jet, so we conclude the \citetalias{Wan_etal1997} setup is much less relevant to astrophysical jets.
}

\subsection{Does the ablation from the angular sides of the groove increase or decrease the velocity of the jet?}

We are now in a position to address whether the ablating plasma from the angular sides of the groove in the target increases or decreases the velocity of the jet.   A key piece of information is that the velocity of the jet is much smaller in the \citetalias{Wan_etal1997}-like experiment in which the target is two Al slabs oriented perpendicular to each other with a gap in between than in the \citetalias{Grava_etal2008} experiment in which the target is an Al slab with a V-shaped groove  in it.  We can now understand the reason why from the answer we obtained to the previous question.  The energy density of the ablating plasma is dominated by its bulk kinetic energy by the time it approaches the mid-plane, except at very small distances ($< 50~\mu$m) from the target.  The component of the momentum of the ablating plasma that is perpendicular to the mid-plane will go into heating the jet, while the component parallel to the mid-plane will add to the velocity of the jet.  However, because the laser intensity is much lower away from the mid-plane, due both to the profile of the laser beam and the slanted angle of the surface of the groove, the {\it specific} internal energy (i.e., the internal energy per gram) generated by the component of the momentum of the ablating plasma when the ablating plasma collides with the jet, and the {\it specific} component of the momentum of the accreting plasma parallel to the mid-plane (i.e., the momentum per unit mass) are both smaller than in the jet flow itself, which is generated by the most intense part of the laser beam illuminating the nearly flat part of the groove near the mid-plane.  This suggests that the entrainment in the jet of the plasma ablating from the sloping sides of the groove will {\it decrease slightly} the velocity of the jet compared to the velocity along the mid-plane of the freely expanding plasma in the case of a slab target.  This expectation is consistent with the results shown in Figure~\ref{fig:vely1}.

\section{Conclusions and Summary}
\label{sec:concl}

We compared the results of FLASH hydrodynamic simulations to previously-published experimental results and HYDRA simulations for the irradiation of a mm-long V-shaped groove cut into an Al target \citetalias{Grava_etal2008}. Importantly, these experiments, conducted at Colorado State University, included soft x-ray interferometric measurements of the electron density in the Al blowoff plasma {\blue that afford a useful} validation test.
We performed these FLASH simulations {\it without} the exact same EOS and opacity models that were used in the previously published HYDRA simulations. Instead we used a commercially available PROPACEOS EOS and opacity model \cite{Macfarlane_etal2006} that included QEOS physics for near-solid-density fluids \cite{QEOS}. 

{\blue We find reasonable agreement between the results of the FLASH simulations and the results of the HYDRA simulations and the experimental measurements presented in \citetalias{Grava_etal2008}.
}

This includes the properties of the underdense Al blowoff plasma, which matters most for the use of these codes in calculating pre-plasma properties as initial conditions for PIC simulations of ultra-intense, short-pulse laser-matter interactions (e.g. \cite{AkliOrban_etal2012}). This result is encouraging for the wider HEDP community since FLASH is a``user" code that is freely available to the academic community. It is also encouraging because FLASH uses a finite-volume AMR scheme that makes it straightforward for the user to configure a simulation to maintain high resolution even in areas where the plasma is expanding rapidly due to laser heating.

{\blue While the results of the FLASH and HYDRA simulations of the \citetalias{Grava_etal2008} experiment are in reasonable agreement, differences do exist. Having accessed the possible sources of the differences, we conclude that future, more detailed code-to-code comparisons involving careful comparisons of the effect of the various components of the FLASH and HYDRA codes seem to be necessary in order to identify with confidence the source or sources of these differences.
}

Having validated the FLASH simulations for the Al target with a V-shaped groove, we used these and other FLASH simulations to better understand the formation and properties of the jet in the experiment. We show that the velocity of the jet is produced primarily by the heating of the target in the relatively flat region of the V-shaped groove at the mid-plane, as in standard slab  targets.  We show that the jet is collimated primarily by the ram pressure of the plasma that ablates from the sloping sides of the groove. Further, we find that the interaction of the plasma ablating from the sloping sides of the groove with the jet produces a complex lateral structure in it. 
{\blue However, the jet that is produced in the experiment is very narrow, the ratio $e_{\rm kin}/e_{\rm int}$ is large along the axis of the jet, and the ratio decreases laterally away from the axis, all of which are properties of astrophysical jets \cite[e.g.][]{Stone_Norman1993}. It is therefore very relevant to astrophysical jets.
}
Finally, we show that the entrainment in the jet of the plasma ablating from the sloping sides of the groove slightly decreases the velocity in the jet compared to the velocity along the mid-plane of the freely expanding plasma in the case of a slab target.

In this work we validated the FLASH code for a specific, previously-published experiment on laser-irradiated Al plasmas. Similarly high-quality interferometric data also exists for C, Cu and Mo  \cite{Filevich_etal2009,Purvis_etal2010}, which can be used in future efforts to validate the HEDP capabilities in FLASH.

\section*{Acknowledgements}
 
Supercomputer allocations for this project included time from the Ohio Supercomputer Center. The FLASH code used in this work was developed in part by the Flash Center for Computational Science at the University of Chicago through funding by DOE NNSA ASC, DOE Office of Science OASCR, and NSF Physics. CO was supported by U.S. Department of Energy contract DE-FG02-05ER54834 (ACE).  DL {\blue and MF were} supported in part at the University of Chicago by the U.S. Department of Energy NNSA ASC through the Argonne Institute for Computing in Science under field work proposal 57789.

\bibliography{ms}

\begin{thebibliography}{43}
\expandafter\ifx\csname natexlab\endcsname\relax\def\natexlab#1{#1}\fi
\expandafter\ifx\csname bibnamefont\endcsname\relax
  \def\bibnamefont#1{#1}\fi
\expandafter\ifx\csname bibfnamefont\endcsname\relax
  \def\bibfnamefont#1{#1}\fi
\expandafter\ifx\csname citenamefont\endcsname\relax
  \def\citenamefont#1{#1}\fi
\expandafter\ifx\csname url\endcsname\relax
  \def\url#1{\texttt{#1}}\fi
\expandafter\ifx\csname urlprefix\endcsname\relax\def\urlprefix{URL }\fi
\providecommand{\bibinfo}[2]{#2}
\providecommand{\eprint}[2][]{\url{#2}}

\bibitem[{Note1()}]{Note1}
Note1, \bibinfo{note}{we use the term ``three-temperature'' (or ``3T'') to
  denote the approximation that electrons and ions move together as a single
  fluid but with two different temperatures, and that this fluid can emit or
  absorb radiation. In the 3T simulations presented in this paper, each cell
  has an electron temperature, an ion temperature, and radiation energy
  densities in a number of photon energy bins.}

\bibitem[{\citenamefont{{Akli} et~al.}(2012)\citenamefont{{Akli}, {Orban},
  {Schumacher}, {Storm}, {Fatenejad}, {Lamb}, and
  {Freeman}}}]{AkliOrban_etal2012}
\bibinfo{author}{\bibfnamefont{K.~U.} \bibnamefont{{Akli}}},
  \bibinfo{author}{\bibfnamefont{C.}~\bibnamefont{{Orban}}},
  \bibinfo{author}{\bibfnamefont{D.}~\bibnamefont{{Schumacher}}},
  \bibinfo{author}{\bibfnamefont{M.}~\bibnamefont{{Storm}}},
  \bibinfo{author}{\bibfnamefont{M.}~\bibnamefont{{Fatenejad}}},
  \bibinfo{author}{\bibfnamefont{D.}~\bibnamefont{{Lamb}}}, \bibnamefont{and}
  \bibinfo{author}{\bibfnamefont{R.~R.} \bibnamefont{{Freeman}}},
  \bibinfo{journal}{Phys. Rev. E.} \textbf{\bibinfo{volume}{86}},
  \bibinfo{eid}{065402} (\bibinfo{year}{2012}).

\bibitem[{\citenamefont{{Boehly} et~al.}(1997)\citenamefont{{Boehly}, {Brown},
  {Craxton}, {Keck}, {Knauer}, {Kelly}, {Kessler}, {Kumpan}, {Loucks},
  {Letzring} et~al.}}]{Boehly_etal1997}
\bibinfo{author}{\bibfnamefont{T.~R.} \bibnamefont{{Boehly}}},
  \bibinfo{author}{\bibfnamefont{D.~L.} \bibnamefont{{Brown}}},
  \bibinfo{author}{\bibfnamefont{R.~S.} \bibnamefont{{Craxton}}},
  \bibinfo{author}{\bibfnamefont{R.~L.} \bibnamefont{{Keck}}},
  \bibinfo{author}{\bibfnamefont{J.~P.} \bibnamefont{{Knauer}}},
  \bibinfo{author}{\bibfnamefont{J.~H.} \bibnamefont{{Kelly}}},
  \bibinfo{author}{\bibfnamefont{T.~J.} \bibnamefont{{Kessler}}},
  \bibinfo{author}{\bibfnamefont{S.~A.} \bibnamefont{{Kumpan}}},
  \bibinfo{author}{\bibfnamefont{S.~J.} \bibnamefont{{Loucks}}},
  \bibinfo{author}{\bibfnamefont{S.~A.} \bibnamefont{{Letzring}}},
  \bibnamefont{et~al.}, \bibinfo{journal}{Optics Communications}
  \textbf{\bibinfo{volume}{133}}, \bibinfo{pages}{495} (\bibinfo{year}{1997}).

\bibitem[{\citenamefont{{Moses} et~al.}(2009)\citenamefont{{Moses}, {Boyd},
  {Remington}, {Keane}, and {Al-Ayat}}}]{Moses_etal2009}
\bibinfo{author}{\bibfnamefont{E.~I.} \bibnamefont{{Moses}}},
  \bibinfo{author}{\bibfnamefont{R.~N.} \bibnamefont{{Boyd}}},
  \bibinfo{author}{\bibfnamefont{B.~A.} \bibnamefont{{Remington}}},
  \bibinfo{author}{\bibfnamefont{C.~J.} \bibnamefont{{Keane}}},
  \bibnamefont{and}
  \bibinfo{author}{\bibfnamefont{R.}~\bibnamefont{{Al-Ayat}}},
  \bibinfo{journal}{Physics of Plasmas} \textbf{\bibinfo{volume}{16}},
  \bibinfo{pages}{041006} (\bibinfo{year}{2009}).

\bibitem[{\citenamefont{{Lindl} et~al.}(2011)\citenamefont{{Lindl}, {Atherton},
  {Amednt}, {Batha}, {Bell}, {Berger}, {Betti}, {Bleuel}, {Boehly}, {Bradley}
  et~al.}}]{Lindl_etal2011}
\bibinfo{author}{\bibfnamefont{J.~D.} \bibnamefont{{Lindl}}},
  \bibinfo{author}{\bibfnamefont{L.~J.} \bibnamefont{{Atherton}}},
  \bibinfo{author}{\bibfnamefont{P.~A.} \bibnamefont{{Amednt}}},
  \bibinfo{author}{\bibfnamefont{S.}~\bibnamefont{{Batha}}},
  \bibinfo{author}{\bibfnamefont{P.}~\bibnamefont{{Bell}}},
  \bibinfo{author}{\bibfnamefont{R.~L.} \bibnamefont{{Berger}}},
  \bibinfo{author}{\bibfnamefont{R.}~\bibnamefont{{Betti}}},
  \bibinfo{author}{\bibfnamefont{D.~L.} \bibnamefont{{Bleuel}}},
  \bibinfo{author}{\bibfnamefont{T.~R.} \bibnamefont{{Boehly}}},
  \bibinfo{author}{\bibfnamefont{D.~K.} \bibnamefont{{Bradley}}},
  \bibnamefont{et~al.}, \bibinfo{journal}{Nuclear Fusion}
  \textbf{\bibinfo{volume}{51}}, \bibinfo{pages}{094024}
  (\bibinfo{year}{2011}).

\bibitem[{\citenamefont{{Rosen} et~al.}(2011)\citenamefont{{Rosen}, {Scott},
  {Hinkel}, {Williams}, {Callahan}, {Town}, {Divol}, {Michel}, {Kruer}, {Suter}
  et~al.}}]{Rosen_etal2011}
\bibinfo{author}{\bibfnamefont{M.~D.} \bibnamefont{{Rosen}}},
  \bibinfo{author}{\bibfnamefont{H.~A.} \bibnamefont{{Scott}}},
  \bibinfo{author}{\bibfnamefont{D.~E.} \bibnamefont{{Hinkel}}},
  \bibinfo{author}{\bibfnamefont{E.~A.} \bibnamefont{{Williams}}},
  \bibinfo{author}{\bibfnamefont{D.~A.} \bibnamefont{{Callahan}}},
  \bibinfo{author}{\bibfnamefont{R.~P.~J.} \bibnamefont{{Town}}},
  \bibinfo{author}{\bibfnamefont{L.}~\bibnamefont{{Divol}}},
  \bibinfo{author}{\bibfnamefont{P.~A.} \bibnamefont{{Michel}}},
  \bibinfo{author}{\bibfnamefont{W.~L.} \bibnamefont{{Kruer}}},
  \bibinfo{author}{\bibfnamefont{L.~J.} \bibnamefont{{Suter}}},
  \bibnamefont{et~al.}, \bibinfo{journal}{High Energy Density Physics}
  \textbf{\bibinfo{volume}{7}}, \bibinfo{pages}{180} (\bibinfo{year}{2011}).

\bibitem[{\citenamefont{{Lamb} and {Marinak}}(2012)}]{LambMarinak2012}
\bibinfo{author}{\bibfnamefont{D.}~\bibnamefont{{Lamb}}} \bibnamefont{and}
  \bibinfo{author}{\bibfnamefont{M.}~\bibnamefont{{Marinak}}},
  \bibinfo{journal}{{Panel Report on Integrated Modeling, Workshop on Science
  of Fusion Ignition on NIF, LLNL-TR-570412}}  (\bibinfo{year}{2012}),
  \urlprefix\url{https://lasers.llnl.gov/workshops/science_of_ignition/}.

\bibitem[{\citenamefont{{Bellei} et~al.}(2013)\citenamefont{{Bellei}, {Amendt},
  {Wilks}, {Haines}, {Casey}, {Li}, {Petrasso}, and {Welch}}}]{Bellei_etal2013}
\bibinfo{author}{\bibfnamefont{C.}~\bibnamefont{{Bellei}}},
  \bibinfo{author}{\bibfnamefont{P.~A.} \bibnamefont{{Amendt}}},
  \bibinfo{author}{\bibfnamefont{S.~C.} \bibnamefont{{Wilks}}},
  \bibinfo{author}{\bibfnamefont{M.~G.} \bibnamefont{{Haines}}},
  \bibinfo{author}{\bibfnamefont{D.~T.} \bibnamefont{{Casey}}},
  \bibinfo{author}{\bibfnamefont{C.~K.} \bibnamefont{{Li}}},
  \bibinfo{author}{\bibfnamefont{R.}~\bibnamefont{{Petrasso}}},
  \bibnamefont{and} \bibinfo{author}{\bibfnamefont{D.~R.}
  \bibnamefont{{Welch}}}, \bibinfo{journal}{Physics of Plasmas}
  \textbf{\bibinfo{volume}{20}}, \bibinfo{pages}{012701}
  (\bibinfo{year}{2013}).

\bibitem[{\citenamefont{Thomas and Kares}(2012)}]{RAGE}
\bibinfo{author}{\bibfnamefont{V.~A.} \bibnamefont{Thomas}} \bibnamefont{and}
  \bibinfo{author}{\bibfnamefont{R.~J.} \bibnamefont{Kares}},
  \bibinfo{journal}{Phys. Rev. Lett.} \textbf{\bibinfo{volume}{109}},
  \bibinfo{pages}{075004} (\bibinfo{year}{2012}),
  \urlprefix\url{http://link.aps.org/doi/10.1103/PhysRevLett.109.075004}.

\bibitem[{\citenamefont{{Marinak} et~al.}(1996)\citenamefont{{Marinak},
  {Tipton}, {Landen}, {Murphy}, {Amendt}, {Haan}, {Hatchett}, {Keane},
  {McEachern}, and {Wallace}}}]{Marinak_etal1996}
\bibinfo{author}{\bibfnamefont{M.~M.} \bibnamefont{{Marinak}}},
  \bibinfo{author}{\bibfnamefont{R.~E.} \bibnamefont{{Tipton}}},
  \bibinfo{author}{\bibfnamefont{O.~L.} \bibnamefont{{Landen}}},
  \bibinfo{author}{\bibfnamefont{T.~J.} \bibnamefont{{Murphy}}},
  \bibinfo{author}{\bibfnamefont{P.}~\bibnamefont{{Amendt}}},
  \bibinfo{author}{\bibfnamefont{S.~W.} \bibnamefont{{Haan}}},
  \bibinfo{author}{\bibfnamefont{S.~P.} \bibnamefont{{Hatchett}}},
  \bibinfo{author}{\bibfnamefont{C.~J.} \bibnamefont{{Keane}}},
  \bibinfo{author}{\bibfnamefont{R.}~\bibnamefont{{McEachern}}},
  \bibnamefont{and}
  \bibinfo{author}{\bibfnamefont{R.}~\bibnamefont{{Wallace}}},
  \bibinfo{journal}{Physics of Plasmas} \textbf{\bibinfo{volume}{3}},
  \bibinfo{pages}{2070} (\bibinfo{year}{1996}).

\bibitem[{\citenamefont{{Marinak} et~al.}(1998)\citenamefont{{Marinak}, {Haan},
  {Dittrich}, {Tipton}, and {Zimmerman}}}]{Marinak_etal1998}
\bibinfo{author}{\bibfnamefont{M.~M.} \bibnamefont{{Marinak}}},
  \bibinfo{author}{\bibfnamefont{S.~W.} \bibnamefont{{Haan}}},
  \bibinfo{author}{\bibfnamefont{T.~R.} \bibnamefont{{Dittrich}}},
  \bibinfo{author}{\bibfnamefont{R.~E.} \bibnamefont{{Tipton}}},
  \bibnamefont{and} \bibinfo{author}{\bibfnamefont{G.~B.}
  \bibnamefont{{Zimmerman}}}, \bibinfo{journal}{Physics of Plasmas}
  \textbf{\bibinfo{volume}{5}}, \bibinfo{pages}{1125} (\bibinfo{year}{1998}).

\bibitem[{\citenamefont{{Marinak} et~al.}(2001)\citenamefont{{Marinak},
  {Kerbel}, {Gentile}, {Jones}, {Munro}, {Pollaine}, {Dittrich}, and
  {Haan}}}]{Marinak_etal2001}
\bibinfo{author}{\bibfnamefont{M.~M.} \bibnamefont{{Marinak}}},
  \bibinfo{author}{\bibfnamefont{G.~D.} \bibnamefont{{Kerbel}}},
  \bibinfo{author}{\bibfnamefont{N.~A.} \bibnamefont{{Gentile}}},
  \bibinfo{author}{\bibfnamefont{O.}~\bibnamefont{{Jones}}},
  \bibinfo{author}{\bibfnamefont{D.}~\bibnamefont{{Munro}}},
  \bibinfo{author}{\bibfnamefont{S.}~\bibnamefont{{Pollaine}}},
  \bibinfo{author}{\bibfnamefont{T.~R.} \bibnamefont{{Dittrich}}},
  \bibnamefont{and} \bibinfo{author}{\bibfnamefont{S.~W.}
  \bibnamefont{{Haan}}}, \bibinfo{journal}{Physics of Plasmas}
  \textbf{\bibinfo{volume}{8}}, \bibinfo{pages}{2275} (\bibinfo{year}{2001}).

\bibitem[{\citenamefont{Harding et~al.}(2009)\citenamefont{Harding, Hansen,
  Hurricane, Drake, Robey, Kuranz, Remington, Bono, Grosskopf, and
  Gillespie}}]{Harding_etal2009}
\bibinfo{author}{\bibfnamefont{E.~C.} \bibnamefont{Harding}},
  \bibinfo{author}{\bibfnamefont{J.~F.} \bibnamefont{Hansen}},
  \bibinfo{author}{\bibfnamefont{O.~A.} \bibnamefont{Hurricane}},
  \bibinfo{author}{\bibfnamefont{R.~P.} \bibnamefont{Drake}},
  \bibinfo{author}{\bibfnamefont{H.~F.} \bibnamefont{Robey}},
  \bibinfo{author}{\bibfnamefont{C.~C.} \bibnamefont{Kuranz}},
  \bibinfo{author}{\bibfnamefont{B.~A.} \bibnamefont{Remington}},
  \bibinfo{author}{\bibfnamefont{M.~J.} \bibnamefont{Bono}},
  \bibinfo{author}{\bibfnamefont{M.~J.} \bibnamefont{Grosskopf}},
  \bibnamefont{and} \bibinfo{author}{\bibfnamefont{R.~S.}
  \bibnamefont{Gillespie}}, \bibinfo{journal}{Phys. Rev. Lett.}
  \textbf{\bibinfo{volume}{103}}, \bibinfo{pages}{045005}
  (\bibinfo{year}{2009}).

\bibitem[{\citenamefont{van~der Holst et~al.}(2012)\citenamefont{van~der Holst,
  Tóth, Sokolov, Daldorff, Powell, and Drake}}]{vanderHolst_etal2012}
\bibinfo{author}{\bibfnamefont{B.}~\bibnamefont{van~der Holst}},
  \bibinfo{author}{\bibfnamefont{G.}~\bibnamefont{Tóth}},
  \bibinfo{author}{\bibfnamefont{I.}~\bibnamefont{Sokolov}},
  \bibinfo{author}{\bibfnamefont{L.}~\bibnamefont{Daldorff}},
  \bibinfo{author}{\bibfnamefont{K.}~\bibnamefont{Powell}}, \bibnamefont{and}
  \bibinfo{author}{\bibfnamefont{R.}~\bibnamefont{Drake}},
  \bibinfo{journal}{High Energy Density Physics} \textbf{\bibinfo{volume}{8}},
  \bibinfo{pages}{161 } (\bibinfo{year}{2012}), ISSN \bibinfo{issn}{1574-1818}.

\bibitem[{\citenamefont{{Taccetti} et~al.}(2012)\citenamefont{{Taccetti},
  {Keiter}, {Lanier}, {Mussack}, {Belle}, and {Magelssen}}}]{Taccetti_etal2012}
\bibinfo{author}{\bibfnamefont{J.~M.} \bibnamefont{{Taccetti}}},
  \bibinfo{author}{\bibfnamefont{P.~A.} \bibnamefont{{Keiter}}},
  \bibinfo{author}{\bibfnamefont{N.}~\bibnamefont{{Lanier}}},
  \bibinfo{author}{\bibfnamefont{K.}~\bibnamefont{{Mussack}}},
  \bibinfo{author}{\bibfnamefont{K.}~\bibnamefont{{Belle}}}, \bibnamefont{and}
  \bibinfo{author}{\bibfnamefont{G.~R.} \bibnamefont{{Magelssen}}},
  \bibinfo{journal}{Review of Scientific Instruments}
  \textbf{\bibinfo{volume}{83}}, \bibinfo{pages}{023506}
  (\bibinfo{year}{2012}).

\bibitem[{\citenamefont{DOE}(2011)}]{NIF}
\bibinfo{author}{\bibnamefont{DOE}}, \bibinfo{journal}{Basic Research
  Directions for User Science at the National Ignition Facility}
  (\bibinfo{year}{2011}).

\bibitem[{\citenamefont{{Keiter} et~al.}(2013)\citenamefont{{Keiter},
  {Mussack}, and {Klein}}}]{Keiter_etal2013}
\bibinfo{author}{\bibfnamefont{P.~A.} \bibnamefont{{Keiter}}},
  \bibinfo{author}{\bibfnamefont{K.}~\bibnamefont{{Mussack}}},
  \bibnamefont{and} \bibinfo{author}{\bibfnamefont{S.~R.}
  \bibnamefont{{Klein}}}, \bibinfo{journal}{High Energy Density Physics}
  \textbf{\bibinfo{volume}{9}}, \bibinfo{pages}{319} (\bibinfo{year}{2013}).

\bibitem[{\citenamefont{{Fryxell} et~al.}(2000)\citenamefont{{Fryxell},
  {Olson}, {Ricker}, {Timmes}, {Zingale}, {Lamb}, {MacNeice}, {Rosner},
  {Truran}, and {Tufo}}}]{Fryxell_etal2000}
\bibinfo{author}{\bibfnamefont{B.}~\bibnamefont{{Fryxell}}},
  \bibinfo{author}{\bibfnamefont{K.}~\bibnamefont{{Olson}}},
  \bibinfo{author}{\bibfnamefont{P.}~\bibnamefont{{Ricker}}},
  \bibinfo{author}{\bibfnamefont{F.~X.} \bibnamefont{{Timmes}}},
  \bibinfo{author}{\bibfnamefont{M.}~\bibnamefont{{Zingale}}},
  \bibinfo{author}{\bibfnamefont{D.~Q.} \bibnamefont{{Lamb}}},
  \bibinfo{author}{\bibfnamefont{P.}~\bibnamefont{{MacNeice}}},
  \bibinfo{author}{\bibfnamefont{R.}~\bibnamefont{{Rosner}}},
  \bibinfo{author}{\bibfnamefont{J.~W.} \bibnamefont{{Truran}}},
  \bibnamefont{and} \bibinfo{author}{\bibfnamefont{H.}~\bibnamefont{{Tufo}}},
  \bibinfo{journal}{\apjs} \textbf{\bibinfo{volume}{131}}, \bibinfo{pages}{273}
  (\bibinfo{year}{2000}).

\bibitem[{\citenamefont{{Dubey} et~al.}(2009)\citenamefont{{Dubey}, {Reid},
  {Weide}, {Antypas}, {Ganapathy}, {Riley}, {Sheeler}, and
  {Siegal}}}]{Dubey_etal2009}
\bibinfo{author}{\bibfnamefont{A.}~\bibnamefont{{Dubey}}},
  \bibinfo{author}{\bibfnamefont{L.~B.} \bibnamefont{{Reid}}},
  \bibinfo{author}{\bibfnamefont{K.}~\bibnamefont{{Weide}}},
  \bibinfo{author}{\bibfnamefont{K.}~\bibnamefont{{Antypas}}},
  \bibinfo{author}{\bibfnamefont{M.~K.} \bibnamefont{{Ganapathy}}},
  \bibinfo{author}{\bibfnamefont{K.}~\bibnamefont{{Riley}}},
  \bibinfo{author}{\bibfnamefont{D.}~\bibnamefont{{Sheeler}}},
  \bibnamefont{and} \bibinfo{author}{\bibfnamefont{A.}~\bibnamefont{{Siegal}}},
  \bibinfo{journal}{ArXiv e-prints}  (\bibinfo{year}{2009}),
  \eprint{0903.4875}.

\bibitem[{\citenamefont{Tzeferacos et~al.}(2015)\citenamefont{Tzeferacos,
  Fatenejad, Flocke, Graziani, Gregori, Lamb, Lee, Meinecke, Scopatz, and
  Weide}}]{Tzeferacos2014}
\bibinfo{author}{\bibfnamefont{P.}~\bibnamefont{Tzeferacos}},
  \bibinfo{author}{\bibfnamefont{M.}~\bibnamefont{Fatenejad}},
  \bibinfo{author}{\bibfnamefont{N.}~\bibnamefont{Flocke}},
  \bibinfo{author}{\bibfnamefont{C.}~\bibnamefont{Graziani}},
  \bibinfo{author}{\bibfnamefont{G.}~\bibnamefont{Gregori}},
  \bibinfo{author}{\bibfnamefont{D.}~\bibnamefont{Lamb}},
  \bibinfo{author}{\bibfnamefont{D.}~\bibnamefont{Lee}},
  \bibinfo{author}{\bibfnamefont{J.}~\bibnamefont{Meinecke}},
  \bibinfo{author}{\bibfnamefont{A.}~\bibnamefont{Scopatz}}, \bibnamefont{and}
  \bibinfo{author}{\bibfnamefont{K.}~\bibnamefont{Weide}},
  \bibinfo{journal}{High Energy Density Physics} \textbf{\bibinfo{volume}{17}},
  \bibinfo{pages}{24 } (\bibinfo{year}{2015}), ISSN \bibinfo{issn}{1574-1818},
  \bibinfo{note}{10th International Conference on High Energy Density
  Laboratory Astrophysics},
  \urlprefix\url{http://www.sciencedirect.com/science/article/pii/S1574181814000779}.

\bibitem[{\citenamefont{{Grava} et~al.}(2008)\citenamefont{{Grava}, {Purvis},
  {Filevich}, {Marconi}, {Rocca}, {Dunn}, {Moon}, and
  {Shlyaptsev}}}]{Grava_etal2008}
\bibinfo{author}{\bibfnamefont{J.}~\bibnamefont{{Grava}}},
  \bibinfo{author}{\bibfnamefont{M.~A.} \bibnamefont{{Purvis}}},
  \bibinfo{author}{\bibfnamefont{J.}~\bibnamefont{{Filevich}}},
  \bibinfo{author}{\bibfnamefont{M.~C.} \bibnamefont{{Marconi}}},
  \bibinfo{author}{\bibfnamefont{J.~J.} \bibnamefont{{Rocca}}},
  \bibinfo{author}{\bibfnamefont{J.}~\bibnamefont{{Dunn}}},
  \bibinfo{author}{\bibfnamefont{S.~J.} \bibnamefont{{Moon}}},
  \bibnamefont{and} \bibinfo{author}{\bibfnamefont{V.~N.}
  \bibnamefont{{Shlyaptsev}}}, \bibinfo{journal}{Phys. Rev. E.}
  \textbf{\bibinfo{volume}{78}}, \bibinfo{eid}{016403} (\bibinfo{year}{2008}).

\bibitem[{\citenamefont{{MacNeice} et~al.}(2000)\citenamefont{{MacNeice},
  {Olson}, {Mobarry}, {de Fainchtein}, and {Packer}}}]{PARAMESH}
\bibinfo{author}{\bibfnamefont{P.}~\bibnamefont{{MacNeice}}},
  \bibinfo{author}{\bibfnamefont{K.~M.} \bibnamefont{{Olson}}},
  \bibinfo{author}{\bibfnamefont{C.}~\bibnamefont{{Mobarry}}},
  \bibinfo{author}{\bibfnamefont{R.}~\bibnamefont{{de Fainchtein}}},
  \bibnamefont{and} \bibinfo{author}{\bibfnamefont{C.}~\bibnamefont{{Packer}}},
  \bibinfo{journal}{Computer Physics Communications}
  \textbf{\bibinfo{volume}{126}}, \bibinfo{pages}{330} (\bibinfo{year}{2000}).

\bibitem[{\citenamefont{{Hirt} et~al.}(1974)\citenamefont{{Hirt}, {Amsden}, and
  {Cook}}}]{Hirt_etal1974}
\bibinfo{author}{\bibfnamefont{C.~W.} \bibnamefont{{Hirt}}},
  \bibinfo{author}{\bibfnamefont{A.~A.} \bibnamefont{{Amsden}}},
  \bibnamefont{and} \bibinfo{author}{\bibfnamefont{J.~L.}
  \bibnamefont{{Cook}}}, \bibinfo{journal}{Journal of Computational Physics}
  \textbf{\bibinfo{volume}{14}}, \bibinfo{pages}{227} (\bibinfo{year}{1974}).

\bibitem[{\citenamefont{{Castor}}(2004)}]{Castor2004}
\bibinfo{author}{\bibfnamefont{J.~I.} \bibnamefont{{Castor}}},
  \emph{\bibinfo{title}{{Radiation Hydrodynamics}}}
  (\bibinfo{publisher}{Cambridge University Press}, \bibinfo{year}{2004}).

\bibitem[{\citenamefont{{Kucharik}}(2006)}]{Kucharik2006}
\bibinfo{author}{\bibfnamefont{M.}~\bibnamefont{{Kucharik}}},
  \bibinfo{journal}{Ph. D. Thesis, Czech Technical University in Prague}
  \textbf{\bibinfo{volume}{27}}, \bibinfo{pages}{1273} (\bibinfo{year}{2006}).

\bibitem[{\citenamefont{Wan et~al.}(1997)\citenamefont{Wan, Barbee, Cauble,
  Celliers, Da~Silva, Moreno, Rambo, Stone, Trebes, and Weber}}]{Wan_etal1997}
\bibinfo{author}{\bibfnamefont{A.~S.} \bibnamefont{Wan}},
  \bibinfo{author}{\bibfnamefont{T.~W.} \bibnamefont{Barbee}},
  \bibinfo{author}{\bibfnamefont{R.}~\bibnamefont{Cauble}},
  \bibinfo{author}{\bibfnamefont{P.}~\bibnamefont{Celliers}},
  \bibinfo{author}{\bibfnamefont{L.~B.} \bibnamefont{Da~Silva}},
  \bibinfo{author}{\bibfnamefont{J.~C.} \bibnamefont{Moreno}},
  \bibinfo{author}{\bibfnamefont{P.~W.} \bibnamefont{Rambo}},
  \bibinfo{author}{\bibfnamefont{G.~F.} \bibnamefont{Stone}},
  \bibinfo{author}{\bibfnamefont{J.~E.} \bibnamefont{Trebes}},
  \bibnamefont{and} \bibinfo{author}{\bibfnamefont{F.}~\bibnamefont{Weber}},
  \bibinfo{journal}{Phys. Rev. E} \textbf{\bibinfo{volume}{55}},
  \bibinfo{pages}{6293} (\bibinfo{year}{1997}),
  \urlprefix\url{http://link.aps.org/doi/10.1103/PhysRevE.55.6293}.

\bibitem[{\citenamefont{{Stone} et~al.}(2000)\citenamefont{{Stone}, {Turner},
  {Estabrook}, {Remington}, {Farley}, {Glendinning}, and
  {Glenzer}}}]{Stone_etal2000}
\bibinfo{author}{\bibfnamefont{J.~M.} \bibnamefont{{Stone}}},
  \bibinfo{author}{\bibfnamefont{N.}~\bibnamefont{{Turner}}},
  \bibinfo{author}{\bibfnamefont{K.}~\bibnamefont{{Estabrook}}},
  \bibinfo{author}{\bibfnamefont{B.}~\bibnamefont{{Remington}}},
  \bibinfo{author}{\bibfnamefont{D.}~\bibnamefont{{Farley}}},
  \bibinfo{author}{\bibfnamefont{S.~G.} \bibnamefont{{Glendinning}}},
  \bibnamefont{and}
  \bibinfo{author}{\bibfnamefont{S.}~\bibnamefont{{Glenzer}}},
  \bibinfo{journal}{\apjs} \textbf{\bibinfo{volume}{127}}, \bibinfo{pages}{497}
  (\bibinfo{year}{2000}).

\bibitem[{\citenamefont{{The Flash Center for Computational
  Science}}(2012)}]{Flash2012}
\bibinfo{author}{\bibnamefont{{The Flash Center for Computational Science}}},
  \bibinfo{journal}{User Guide Version-4.0-beta}  (\bibinfo{year}{2012}).

\bibitem[{\citenamefont{{Lee} and {More}}(1984)}]{LeeMore1984}
\bibinfo{author}{\bibfnamefont{Y.~T.} \bibnamefont{{Lee}}} \bibnamefont{and}
  \bibinfo{author}{\bibfnamefont{R.~M.} \bibnamefont{{More}}},
  \bibinfo{journal}{Physics of Fluids} \textbf{\bibinfo{volume}{27}},
  \bibinfo{pages}{1273} (\bibinfo{year}{1984}).

\bibitem[{fla()}]{flash_versions}
\emph{\bibinfo{title}{Flash code user guides}},
  \bibinfo{howpublished}{\url{http://flash.uchicago.edu/site/flashcode/user_support/}}.

\bibitem[{\citenamefont{{Macfarlane} et~al.}(2006)\citenamefont{{Macfarlane},
  {Golovkin}, and {Woodruff}}}]{Macfarlane_etal2006}
\bibinfo{author}{\bibfnamefont{J.~J.} \bibnamefont{{Macfarlane}}},
  \bibinfo{author}{\bibfnamefont{I.~E.} \bibnamefont{{Golovkin}}},
  \bibnamefont{and} \bibinfo{author}{\bibfnamefont{P.~R.}
  \bibnamefont{{Woodruff}}}, \bibinfo{journal}{JSQRT}
  \textbf{\bibinfo{volume}{99}}, \bibinfo{pages}{381} (\bibinfo{year}{2006}).

\bibitem[{\citenamefont{{More} et~al.}(1988)\citenamefont{{More}, {Warren},
  {Young}, and {Zimmerman}}}]{QEOS}
\bibinfo{author}{\bibfnamefont{R.~M.} \bibnamefont{{More}}},
  \bibinfo{author}{\bibfnamefont{K.~H.} \bibnamefont{{Warren}}},
  \bibinfo{author}{\bibfnamefont{D.~A.} \bibnamefont{{Young}}},
  \bibnamefont{and} \bibinfo{author}{\bibfnamefont{G.~B.}
  \bibnamefont{{Zimmerman}}}, \bibinfo{journal}{Physics of Fluids}
  \textbf{\bibinfo{volume}{31}}, \bibinfo{pages}{3059} (\bibinfo{year}{1988}).

\bibitem[{\citenamefont{{Seaton} et~al.}(1994)\citenamefont{{Seaton}, {Yan},
  {Mihalas}, and {Pradhan}}}]{Seaton_etal1994}
\bibinfo{author}{\bibfnamefont{M.~J.} \bibnamefont{{Seaton}}},
  \bibinfo{author}{\bibfnamefont{Y.}~\bibnamefont{{Yan}}},
  \bibinfo{author}{\bibfnamefont{D.}~\bibnamefont{{Mihalas}}},
  \bibnamefont{and} \bibinfo{author}{\bibfnamefont{A.~K.}
  \bibnamefont{{Pradhan}}}, \bibinfo{journal}{\mnras}
  \textbf{\bibinfo{volume}{266}}, \bibinfo{pages}{805} (\bibinfo{year}{1994}).

\bibitem[{\citenamefont{{Mendoza} et~al.}(2007)\citenamefont{{Mendoza},
  {Seaton}, {Buerger}, {Bellor{\'{\i}}n}, {Mel{\'e}ndez}, {Gonz{\'a}lez},
  {Rodr{\'{\i}}guez}, {Delahaye}, {Palacios}, {Pradhan}
  et~al.}}]{Mendoza_etal2007}
\bibinfo{author}{\bibfnamefont{C.}~\bibnamefont{{Mendoza}}},
  \bibinfo{author}{\bibfnamefont{M.~J.} \bibnamefont{{Seaton}}},
  \bibinfo{author}{\bibfnamefont{P.}~\bibnamefont{{Buerger}}},
  \bibinfo{author}{\bibfnamefont{A.}~\bibnamefont{{Bellor{\'{\i}}n}}},
  \bibinfo{author}{\bibfnamefont{M.}~\bibnamefont{{Mel{\'e}ndez}}},
  \bibinfo{author}{\bibfnamefont{J.}~\bibnamefont{{Gonz{\'a}lez}}},
  \bibinfo{author}{\bibfnamefont{L.~S.} \bibnamefont{{Rodr{\'{\i}}guez}}},
  \bibinfo{author}{\bibfnamefont{F.}~\bibnamefont{{Delahaye}}},
  \bibinfo{author}{\bibfnamefont{E.}~\bibnamefont{{Palacios}}},
  \bibinfo{author}{\bibfnamefont{A.~K.} \bibnamefont{{Pradhan}}},
  \bibnamefont{et~al.}, \bibinfo{journal}{\mnras}
  \textbf{\bibinfo{volume}{378}}, \bibinfo{pages}{1031} (\bibinfo{year}{2007}),
  \eprint{0704.1583}.

\bibitem[{\citenamefont{Kaiser}(2000)}]{Kaiser2000}
\bibinfo{author}{\bibfnamefont{T.~B.} \bibnamefont{Kaiser}},
  \bibinfo{journal}{Phys. Rev. E} \textbf{\bibinfo{volume}{61}},
  \bibinfo{pages}{895} (\bibinfo{year}{2000}),
  \urlprefix\url{http://link.aps.org/doi/10.1103/PhysRevE.61.895}.

\bibitem[{\citenamefont{{Shafranov}}(1957)}]{Shafranov1957}
\bibinfo{author}{\bibfnamefont{V.~D.} \bibnamefont{{Shafranov}}},
  \bibinfo{journal}{Sov. Phys. JETP 5, 1183}  (\bibinfo{year}{1957}).

\bibitem[{\citenamefont{{Mihalas} and {Mihalas}}(1984)}]{MihalasMihalas1984}
\bibinfo{author}{\bibfnamefont{D.}~\bibnamefont{{Mihalas}}} \bibnamefont{and}
  \bibinfo{author}{\bibfnamefont{B.~W.} \bibnamefont{{Mihalas}}},
  \emph{\bibinfo{title}{{Foundations of radiation hydrodynamics}}}
  (\bibinfo{year}{1984}).

\bibitem[{\citenamefont{{Fatenejad} et~al.}(2011)\citenamefont{{Fatenejad},
  {Fryer}, {Fryxell}, {Lamb}, {Myra}, and {Wohlbier}}}]{Fatenejad_etal2011}
\bibinfo{author}{\bibfnamefont{M.}~\bibnamefont{{Fatenejad}}},
  \bibinfo{author}{\bibfnamefont{C.}~\bibnamefont{{Fryer}}},
  \bibinfo{author}{\bibfnamefont{B.}~\bibnamefont{{Fryxell}}},
  \bibinfo{author}{\bibfnamefont{D.}~\bibnamefont{{Lamb}}},
  \bibinfo{author}{\bibfnamefont{E.}~\bibnamefont{{Myra}}}, \bibnamefont{and}
  \bibinfo{author}{\bibfnamefont{J.}~\bibnamefont{{Wohlbier}}}, in
  \emph{\bibinfo{booktitle}{APS Meeting Abstracts}} (\bibinfo{year}{2011}), p.
  \bibinfo{pages}{8004}.

\bibitem[{\citenamefont{{Filevich} et~al.}(2009)\citenamefont{{Filevich},
  {Purvis}, {Grava}, {Ryan}, {Dunn}, {Moon}, {Shlyaptsev}, and
  {Rocca}}}]{Filevich_etal2009}
\bibinfo{author}{\bibfnamefont{J.}~\bibnamefont{{Filevich}}},
  \bibinfo{author}{\bibfnamefont{M.}~\bibnamefont{{Purvis}}},
  \bibinfo{author}{\bibfnamefont{J.}~\bibnamefont{{Grava}}},
  \bibinfo{author}{\bibfnamefont{D.~P.} \bibnamefont{{Ryan}}},
  \bibinfo{author}{\bibfnamefont{J.}~\bibnamefont{{Dunn}}},
  \bibinfo{author}{\bibfnamefont{S.~J.} \bibnamefont{{Moon}}},
  \bibinfo{author}{\bibfnamefont{V.~N.} \bibnamefont{{Shlyaptsev}}},
  \bibnamefont{and} \bibinfo{author}{\bibfnamefont{J.~J.}
  \bibnamefont{{Rocca}}}, \bibinfo{journal}{High Energy Density Physics}
  \textbf{\bibinfo{volume}{5}}, \bibinfo{pages}{276} (\bibinfo{year}{2009}).

\bibitem[{\citenamefont{{Purvis} et~al.}(2010)\citenamefont{{Purvis}, {Grava},
  {Filevich}, {Ryan}, {Moon}, {Dunn}, {Shlyaptsev}, and
  {Rocca}}}]{Purvis_etal2010}
\bibinfo{author}{\bibfnamefont{M.~A.} \bibnamefont{{Purvis}}},
  \bibinfo{author}{\bibfnamefont{J.}~\bibnamefont{{Grava}}},
  \bibinfo{author}{\bibfnamefont{J.}~\bibnamefont{{Filevich}}},
  \bibinfo{author}{\bibfnamefont{D.~P.} \bibnamefont{{Ryan}}},
  \bibinfo{author}{\bibfnamefont{S.~J.} \bibnamefont{{Moon}}},
  \bibinfo{author}{\bibfnamefont{J.}~\bibnamefont{{Dunn}}},
  \bibinfo{author}{\bibfnamefont{V.~N.} \bibnamefont{{Shlyaptsev}}},
  \bibnamefont{and} \bibinfo{author}{\bibfnamefont{J.~J.}
  \bibnamefont{{Rocca}}}, \bibinfo{journal}{Phys. Rev. E.}
  \textbf{\bibinfo{volume}{81}}, \bibinfo{eid}{036408} (\bibinfo{year}{2010}).

\bibitem[{\citenamefont{Zimmerman and Kruer}(1975)}]{LASNEX}
\bibinfo{author}{\bibfnamefont{G.~B.} \bibnamefont{Zimmerman}}
  \bibnamefont{and} \bibinfo{author}{\bibfnamefont{W.~L.} \bibnamefont{Kruer}},
  \bibinfo{journal}{Comments Plasma Phys. Controlled Fusion}
  \textbf{\bibinfo{volume}{2}}, \bibinfo{pages}{51} (\bibinfo{year}{1975}).

\bibitem[{\citenamefont{Gao et~al.}(2019)\citenamefont{Gao, Liang, Lu, Follet,
  Sio, Tzeferacos, Froula, Birkel, Li, Lamb et~al.}}]{Gao_etal2019}
\bibinfo{author}{\bibfnamefont{L.}~\bibnamefont{Gao}},
  \bibinfo{author}{\bibfnamefont{E.}~\bibnamefont{Liang}},
  \bibinfo{author}{\bibfnamefont{Y.}~\bibnamefont{Lu}},
  \bibinfo{author}{\bibfnamefont{R.~K.} \bibnamefont{Follet}},
  \bibinfo{author}{\bibfnamefont{H.}~\bibnamefont{Sio}},
  \bibinfo{author}{\bibfnamefont{P.}~\bibnamefont{Tzeferacos}},
  \bibinfo{author}{\bibfnamefont{D.~H.} \bibnamefont{Froula}},
  \bibinfo{author}{\bibfnamefont{A.}~\bibnamefont{Birkel}},
  \bibinfo{author}{\bibfnamefont{C.~K.} \bibnamefont{Li}},
  \bibinfo{author}{\bibfnamefont{D.}~\bibnamefont{Lamb}}, \bibnamefont{et~al.},
  \bibinfo{journal}{The Astrophysical Journal} \textbf{\bibinfo{volume}{873}},
  \bibinfo{pages}{L11} (\bibinfo{year}{2019}),
  \urlprefix\url{https://doi.org/10.3847/2041-8213/ab07bd}.

\bibitem[{\citenamefont{{Stone} and {Norman}}(1993)}]{Stone_Norman1993}
\bibinfo{author}{\bibfnamefont{J.~M.} \bibnamefont{{Stone}}} \bibnamefont{and}
  \bibinfo{author}{\bibfnamefont{M.~L.} \bibnamefont{{Norman}}},
  \bibinfo{journal}{\apj} \textbf{\bibinfo{volume}{413}}, \bibinfo{pages}{198}
  (\bibinfo{year}{1993}).

\end{thebibliography}
\bibliographystyle{apsrev}

\end{document}